\documentclass[twocolumn,aps,epsfig,nofootinbib]{revtex4}

\usepackage{graphicx}
\usepackage{epstopdf}
\usepackage{latexsym}
\usepackage{amssymb}
\usepackage{amsmath}
\usepackage{color}
\usepackage{mathrsfs}

\usepackage[center]{subfigure}

\begin{document}

 \newcommand{\bq}{\begin{equation}}
 \newcommand{\eq}{\end{equation}}
 \newcommand{\bqn}{\begin{eqnarray}}
 \newcommand{\eqn}{\end{eqnarray}}
 \newcommand{\nb}{\nonumber}
 \newcommand{\lb}{\label}
\newcommand{\PRL}{Phys. Rev. Lett.}
\newcommand{\PL}{Phys. Lett.}
\newcommand{\PR}{Phys. Rev.}
\newcommand{\PRD}{Phys. Rev. D.}
\newcommand{\CQG}{Class. Quantum Grav.}
\newcommand{\JCAP}{J. Cosmol. Astropart. Phys.}
\newcommand{\JHEP}{J. High. Energy. Phys.}

\title{Gravitational quantum effects on  power spectra and spectral indices  with higher-order corrections}

\author{Tao Zhu, Anzhong Wang \footnote{The corresponding author}}

 \affiliation{ Institute for Advanced Physics $\&$ Mathematics, Zhejiang University of Technology, Hangzhou, 310032, China\\
 GCAP-CASPER, Physics Department, Baylor University, Waco, TX 76798-7316, USA}

 \author{Gerald Cleaver}

 \affiliation{EUCOS-CASPER, Physics Department, Baylor University, Waco, TX 76798-7316, USA}

 \author{Klaus Kirsten and Qin Sheng}

 \affiliation{GCAP-CASPER, Mathematics Department, Baylor University, Waco, TX 76798-7328, USA}

\date{\today}

\begin{abstract}

The uniform asymptotic approximation method provides a powerful, systematically-improved, and error-controlled approach
to construct accurate analytical approximate solutions of mode functions of  perturbations of the Friedmann-Robertson-Walker
universe, designed especially for the cases where the relativistic linear dispersion relation is modified after gravitational quantum effects are
taken into account. These include models  from string/M-Theory,  loop quantum cosmology and Ho\v{r}ava-Lifshitz quantum gravity.
In this paper, we extend our previous studies  of the first-order approximations to high orders  for  the cases where the modified
dispersion relation (linear or nonlinear) has only one-turning point (or zero). We obtain the general expressions for the power
spectra and spectral indices  of both scalar and tensor perturbations up  to the third-order, at which the error bounds  are
$\lesssim 0.15\%$. As an application of these formulas, we calculate the power spectra and spectral indices  in the slow-roll
inflation with a nonlinear power-law dispersion relation. To check the consistency of our formulas, we further restrict ourselves to
the relativistic case,  and calculate the corresponding power spectra, spectral indices and runnings  to the second-order. Then, we
compare our  results  with the ones obtained by the Green function method, and show explicitly that the results  obtained by these
two different methods are consistent within the allowed errors.

\end{abstract}



\maketitle

\section{Introduction}
\renewcommand{\theequation}{1.\arabic{equation}} \setcounter{equation}{0}

The cosmological inflation not only solves most problems of the standard big bang cosmology, but also provides the simplest and most elegant mechanism to produce the primordial density perturbations and primordial gravitational waves (PGWs) \cite{Guth,InfGR}. The former grows to produce the observed large-scale structure of the universe, and meanwhile creates the cosmic microwave background (CMB) temperature anisotropy, which was already detected by various CMB observations \cite{COBE,WMAP,PLANCK} and galaxy surveys \cite{GSs}. PGWs, on the other hand, produce not only a temperature anisotropy, but also a distinguishable signature in CMB polarization - the B mode polarization, which has been observed recently by the ground based BICEP2 experiment \cite{BICEP2}. These observations  have unprecedented  precisions in the  measurement of the power spectra and spectral indices, and together with the  forthcoming ones,   provide unique  opportunities for us to get deep insight on the physics of  the very early universe. More important, they could provide an unique  window  to explore gravitational quantum effects, which otherwise cannot be studied in the near future by any  man-made lab experiments.

On the other hand, as is well known, the inflationary scenario is conceptually incomplete in serval respects \cite{Brandenberger1999,Martin2001}. For example, in most of the inflation models, the energy scale of quantum fluctuations, which relate to the present observations, were not far from the Planck scale at the beginning of inflation. Thus, questions immediately arise as to whether the usual predictions of the scenario in the framework of semi-classical approximations still remain robust due to the ignorance of gravitational quantum physics at such high energy and, more interestingly, whether they could leave imprints for future observations. Further problems of inflationary scenario include the existence of the initial singularities  \cite{sing}, and the particular initial conditions \cite{Burgess}.
Yet,  a large tensor-to-scalar ratio $r\sim 0.16$, measured recently by   BICEP2 \cite{BICEP2}, leads to the so-called Planckian excursion of the inflaton field, which   makes  the effective theory of inflation questionable \cite{ZW14}.

All the problems mentioned above are closely related to the high energy physics that the usual classical general relativity (GR) and effective field theory of inflation are known to break down. It is widely expected that new physics in this regime --- quantum theory of gravity, will provide a complete description of the early universe. Although such a theory  has not been properly formulated, gravitational quantum effects on inflation have already  been studied extensively  by various approaches. See, for example,  \cite{Brandenberger1999,Martin2001,Ashtekar,Burgess,ZW14,eff,Brandenberger2013CQG}, and references therein. In most of these considerations, both the  scalar and tensor perturbations produced during the inflationary epoch are governed by the equation
\bqn
\lb{modeeom}
\mu_k''(\eta)+\left(\omega_k^2(\eta)-\frac{z''(\eta)}{z(\eta)}\right)\mu_k(\eta)=0,
\eqn
where $\mu_k(\eta)$ denotes the  mode function, a prime the differentiation with respect to the conformal time $\eta$. $k$ is the comoving wavenumber, and $z(\eta)$ depends on the background and the types of perturbations (scalar and tensor).  The modified dispersion relation $\omega_k^2(\eta)$ depends on both the specified quantum gravity models and scenarios of inflation. For example, in the Ho\v{r}ava-Lifshitz quantum gravity \cite{Horava}, $\omega_k^2(\eta)$ takes the form \cite{HL}
\bqn\lb{disp}
\omega_k^2(\eta)=k^2\left[1-\hat b_1 \left(\frac{k}{aM_*}\right)^2+\hat b_2 \left(\frac{k}{aM_*}\right)^4\right],
\eqn
where $M_*$ is the relevant energy scale of the trans-Planckian physics, $a=a(\eta)$ is the scalar factor of the background universe, $\hat b_1$ and $\hat b_2$ are dimensionless constants. When $\hat b_1=0=\hat b_2$, it reduces to that of GR,
\bq
\lb{1.3}
\omega_k^2(\eta)=k^2.
\eq

The dispersion relation is also  modified  in   models of string/M-theory. For example, in the DBI inflation \cite{DBI} it takes the form
\bqn
\omega_k^2(\eta)= c_s^2(\eta) k^2,
\eqn
where $c_s(\eta)$ is the effective sound speed. Interestingly, $c_s^2(\eta)$ could be very close to zero in the far UV regime \cite{Martin_K1,Martin_K2}.

In  the framework of loop quantum  cosmology,  $\omega_k^2(\eta)$ is also modified.   For example, with the holonomy corrections \cite{holonomy,loop_corrections}, $\omega_k^2(\eta)$
becomes
\bqn
\omega_k^2(\eta) = \left(1- \frac{2 \rho(\eta)}{\rho_c}\right) k^2,
\eqn
where $\rho(\eta)$ denotes the energy density of the universe, and $\rho_c$  the critical energy density at which a big bounce occurs, whereby  the classical initial singularity of the universe
is removed. On the other hand,  the inverse-volume corrections \cite{loop_corrections,Bojowald2008b} lead to
\bqn
\omega_k^2(\eta)= \begin{cases}
s^2(\eta)k^2,  & \mbox{ scalar},\cr
(1+2 \hat \alpha_0 \delta_{\text{PL}}) k^2, & \mbox{tensor},\cr
\end{cases}
\eqn
with
\bqn
s^2(\eta)=1+\left[\frac{\sigma \nu_0}{3} \left(\frac{\sigma}{6}+1\right)+\frac{\hat \alpha_0}{2} \left(5-\frac{\sigma}{3}\right)\right]\delta_{PL}(\eta),\nb
\eqn
where $\hat \alpha_0,\;\nu_0,\;\text{and}\; \sigma$ encode the specific features of the model.

In addition to the above mentioned models,  mode functions with other modified dispersion relations have also been studied  phenomenologically to mimic  quantum gravitational effects
in the very early universe \cite{dissipative,lorentzbrane}.

To study these  quantum effects, a critical step is  to solve Eq.(\ref{modeeom}) analytically for the mode function $\mu_k(\eta)$,   and then extract information from it,
including the power spectra, spectral indices and runnings.  Such studies are very challenging, as the problem becomes mathematically very much involved, and meanwhile the treatment
needs to be very accurate, in order to match with the  observational  data of current and forthcoming experiments. Currently,  in most of the analytical treatments  of the mode functions with modified
dispersion relations the corresponding errors are unknown, and   frequently    are by far beyond  the required accuracy by observations. This yields a big gap between the treatments
of  these rich phenomenological models of the very early universe and the high precision observational data.

The attempt to close this gap was initiated by Habib et al  \cite{uniformPRL}, in which  the {\em uniform asymptotic approximation} was first introduced to inflationary cosmology. However, they considered  only the relativistic case, in which  the dispersion relation is the standard linear one given by  Eq.(\ref{1.3}).
In addition,  the mode function was constructed only to the first-order approximation, for which the error bounds in general are $\lesssim 15\%$ \footnote{With further improvement,
the error bounds can be dramatically lowered in some particular cases \cite{uniformPRD}.}.  Clearly, this is not accurate enough to match with the accuracy of the observational data.
Moreover, their results  cannot be applied to the cases with   nonlinear dispersion relations, mentioned above.

In  order to account for the   quantum effects,   recently we generalized the  uniform asymptotic approximation method of Habib et al  \cite{uniformPRL,uniformPRD} to the cases with non-linear dispersion relations, where multiple and high-order turning points are allowed \cite{Zhu2}.  From such obtained  approximate solutions of the mode functions,  we  calculated the power spectra and the corresponding spectral indices of scalar and tensor perturbations to the first-order approximation. However, as pointed out above, in general the error bounds are only $\lesssim 15\%$, although in some cases further improvement can be done,
as in the relativistic case \cite{uniformPRD}.

To match with the accuracy of the current and forthcoming  observations, we  need at least to consider the slow-roll approximation to the second-order, which requires considerations of the high-order approximations even in the framework of the uniform asymptotic approximations.  This is precisely what we are going to do in this paper.
In particular, by working out explicitly  the high-order uniform asymptotic approximations, we obtain the accurate analytical solutions of the mode function at an arbitrary high-order   for the case where only one single turning point exists. We  construct explicitly the error bounds associated with the approximations order by order. With the accurate analytical solutions of the mode functions, we obtain  the general expressions of the power spectra and spectral indices up to the third-order. 

As an application of the developed formulas, we calculate explicitly the power spectra and spectral indices of scalar and tensor perturbations with the modified dispersion relation given by Eq.(\ref{disp}) in the slow-roll inflation approximation. To test the consistency of our formulas, we further restrict ourselves to  the relativistic case ($\hat b_1 = \hat b_2 = 0$),  in which  the power spectra, spectral indices, and the corresponding runnings of spectral indices have been calculated to the second-order by using the Green function method \cite{green1,green2} \footnote{To the second-order, the power spectra and spectral indices in GR
were also obtained by the improved WKB method \cite{WKB}. In addition, K-inflationary power spectra at the second-order of the slow-roll parameters were  calculated by using the first-order approximations of the
uniform asymptotic approximation method \cite{Martin_K1,Martin_K2}.}. We compare the results obtained by these two different methods,   and show that they are essentially the same within the allowed errors.
In this case, we also compare our results order by order with the numerical (exact) results for the evolution of the mode functions, and the results are illustrated in Fig.\ref{fig1}, from which one can see that
our analytical results to the second- and third-order approximations are extremely closed to the exact results.   

Specifically, the paper is organized as follows. In Sec. II, we present a brief review of the uniform asymptotic approximation method. In the case where there is only one turning point, the mode function is  expanded in terms of $1/\lambda$ to an arbitrarily high order, in which the error bounds are given in each of the orders.   In Sec. III we obtain the general expressions of the power spectra and spectral indices to the third-order  from the analytical approximate mode function given in Sec. II, after first matching it with the initial conditions.  In Sec. IV, we apply the general expressions  obtained in Sec. III to the inflationary model with the nonlinear power-law dispersion relation (\ref{disp}), and calculate the  power spectra,
and  spectral indices in the slow-roll inflation. In Sec. V, we consider the GR limit of the power spectra and spectral indices, and also calculate the runnings of the spectral indices and the tensor-to-scalar ratio. Furthermore, we compare the power spectra and spectral indices with those obtained by the Green function method, and show that the two sets of expressions are essentially identical within the allowed errors. Our main conclusions are  summarized  in Sec. VI. Four appendices, A-D,  are also included, in which various detailed mathematical calculations  are presented.

\section{The uniform asymptotic approximations}
\renewcommand{\theequation}{2.\arabic{equation}} \setcounter{equation}{0}

In this section, we present a brief  introduction to the {\em uniform asymptotic approximation method} for the cases where the dispersion relation has only a single turning point.

Following refs. \cite{uniformPRL,Olver1974,ZWCKS,Zhu2}, by introducing a dimensionless variable $y=-k\eta$, let us first write the equation of the mode function in the  form
\bqn\lb{eom}
\frac{d^2\mu_k(y)}{dy^2}=\left[\lambda^2\hat g(y)+q(y)\right]\mu_k(y).
\eqn
In the above the parameter $\lambda$ is used to trace the order of the uniform approximations, and $\lambda^2\hat g(y)=g(y)$. Usually $\lambda$ is supposed to be large, and it also can be absorbed into $g(y)$ thus when we turn to determine the final results, we can set $\lambda=1$ for the sake of simplification. From Eq.(\ref{modeeom}) we find that
\bqn
\lambda^2\hat g(y)+q(y) \equiv - \frac{1}{k^2} \left(\omega^2_k(\eta)-\frac{z''(\eta)}{z(\eta)}\right).
\eqn
In most of the cases, $\hat g(y)$ and $q(y)$ have two poles (singularities): one is at $y=0^+$ and the other is at $y=+\infty$. As we discussed in \cite{Zhu2} (see also \cite{uniformPRL,Olver1974}), if these two poles are both second-order or higher, one has to choose
\bq
\lb{qF}
q(y)=- \frac{1}{4y^2}, 
\eq
for the convergence of the error control functions. In this paper we shall restrict our discussions to this choice. In addition, the function $\hat g(y)$ can vanish at various points, which are called turning points or zeros, and the approximate solution of the mode function $\mu_k(y)$ depends on the behavior of $\hat g(y)$ and $q(y)$ near the turning points.

To proceed further, let us first introduce the Liouville transformations with two new variables $U(\xi)$ and $\xi$ via the relations,
\bqn
\lb{Olver trans}
U(\xi)&=& \chi^{1/4} \mu_k(y),\;\;\; \xi'^2 =  \frac{|g(y)|}{f^{(1)}(\xi)^2},
\eqn
where $ \chi \equiv \xi'^2,\; \xi'=d\xi/dy$, and
\bqn
\lb{OlverTransB}
f(\xi)&=& \int^y \sqrt{|g(y)|} dy,\;\;\;  f^{(1)}(\xi)=\frac{df(\xi)}{d\xi}.
\eqn
Note that $\chi$ must be regular and not vanish in the intervals of interest. Consequently, $f(\xi)$ must be chosen so that
$f^{(1)}(\xi)$ has zeros and singularities of the same type   as that of $g(y)$. As shown below, such requirement plays an
essential role in determining the approximate solutions.
 In terms of $U$ and $\xi$, Eq.(\ref{eom}) takes the form
\bqn\lb{eomU}
\frac{d^2 U}{d\xi^2}&=&\left[\pm f^{(1)}(\xi)^2+\psi(\xi)\right]U,
\eqn
where
\bqn\lb{psi}
\psi(\xi)=\frac{q(y)}{\chi}-\chi^{-3/4} \frac{d^2(\chi^{-1/4})}{dy^2},
\eqn
and the signs ``$\pm$" correspond to $g(y)>0$ and $g(y)<0$, respectively. Considering $\psi(\xi) =0$ as the first-order approximation,
one can choose $f^{(1)}(\xi)$ so that the first-order approximation can be as close
to the exact solution as possible with the guidelines of the error functions constructed below, and then  solve it in terms of known functions.
Clearly, such a choice  sensitively  depends on the behavior of the functions $g(y)$ and $q(y)$ near  the poles and turning points.

In  this paper,   we  consider only the case in which $\hat g(y)$ has only one single turning point $\bar{y}_0$
(for $\hat g(y)$ having several different turning points or one multiple-turning point,
see \cite{Zhu2}), i.e., $\hat g(\bar y_0)=0$.
In  this case we can choose
\bqn
f^{(1)}(\xi)=\pm \xi,
\eqn
here $\xi=\xi(y)$ is a monotone decreasing function, and $\pm$ correspond to $\hat g(y)\geq 0$ and $\hat g(y) \leq 0$, respectively.
Following Olver  \cite{Olver1974}, the general solution of Eq.(\ref{eomU}) can be written as
\bqn\lb{appro}
U(\xi)&=&\alpha_0 \Bigg[\text{Ai}(\lambda^{2/3} \xi) \sum_{s=0}^{n} \frac{A_s(\xi)}{\lambda^{2s}}\nb\\
&&~~~~~~~+\frac{\text{Ai}'(\lambda^{2/3}\xi)}{\lambda^{4/3}} \sum_{s=0}^{n-1} \frac{B_s(\xi)}{\lambda^{2s}}+\epsilon_3^{(2n+1)}\Bigg]\nb\\
&&+\beta_0 \Bigg[\text{Bi}(\lambda^{2/3} \xi) \sum_{s=0}^{n} \frac{A_s(\xi)}{\lambda^{2s}}\nb\\
&&~~~~~~~~+\frac{\text{Bi}'(\lambda^{2/3}\xi)}{\lambda^{4/3}} \sum_{s=0}^{n-1} \frac{B_s(\xi)}{\lambda^{2s}}+\epsilon_4^{(2n+1)}\Bigg],\nb\\
\eqn
where $\text{Ai}(x)$ and $\text{Bi}(x)$ represent the Airy  functions, $\epsilon_3^{(2n+1)}$ and $\epsilon_{4}^{(2n+1)}$ are errors of the approximate solution, and
\bqn\lb{AB}
&& A_0(\xi)=1,\;\;\nb\\
&&B_s(\xi)=\frac{\pm 1}{2 (\pm \xi)^{1/2}}\int_0^\xi \{\psi(v) A_s(v)-A''_s(v)\}\frac{dv}{(\pm v)^{1/2}},\nb\\
&&A_{s+1}(\xi)=-\frac{1}{2} B'_s(\xi)+\frac{1}{2} \int \psi(v) B_s(v) dv,\nb\\
\eqn
where $\pm$ correspond to $\xi \geq 0$ and $\xi\leq 0$, respectively. The error bounds of $\epsilon_3^{(2n+1)}$ and $\epsilon_4^{(2n+1)}$ can be expressed as
\bqn\lb{error}
&&\frac{\epsilon_{3}^{(2n+1)}}{M(\lambda^{2/3} \xi)}, \;\;\frac{\partial \epsilon_{3}^{(2n+1)}/\partial\xi}{\lambda^{2/3} N(\lambda^{2/3}\xi)}\nb\\
&& ~~~~~\leq 2 E^{-1}(\lambda^{2/3}\xi)  \exp{\left[\frac{2\kappa_0 \mathscr{V}_{\alpha,\xi}(|\xi^{1/2}|B_0)}{\lambda}\right]} \nb\\
&&~~~~~~~~~~~\times \frac{\mathscr{V}_{\alpha,\xi}(|\xi^{1/2}|B_n)}{\lambda^{2n+1}},\nb\\
&&\frac{\epsilon_{4}^{(2n+1)}}{M(\lambda^{2/3} \xi)}, \;\;\frac{\partial \epsilon_{4}^{(2n+1)}/\partial\xi}{\lambda^{2/3} N(\lambda^{2/3}\xi)} \nb\\
&&~~~~~\leq 2 E(\lambda^{2/3}\xi) \exp{\left[\frac{2\kappa_0 \mathscr{V}_{\xi,\beta}(|\xi^{1/2}|B_0)}{\lambda}\right]} \nb\\
&&~~~~~~~~~~~\times\frac{\mathscr{V}_{\xi,\beta}(|\xi^{1/2}|B_n)}{\lambda^{2n+1}},
\eqn
where the definitions of $M(x)$, $N(x)$, $\kappa_0$, and $\mathscr{V}_{a,b}(x)$ can be found in \cite{Zhu2}.

Having obtained the approximate solution of the mode function $\mu_k(y)$, let us compare it with the numerical solution order by order. For the sake of simplification, we consider a linear dispersion relation $\omega_k^2(\eta)=k^2$ and a de-Sitter background, i.e, $\nu=3/2$. In the top panel of Fig.\ref{fig1}, we present the numerical (exact) solution, the approximate solution of the first-,  second-, and third-orderapproximations, respectively. From this figure one can see clearly that our analytical solutions to the    second- and third-order  approximations are extremely closed to the exact ones. In practice, our analytical  solution of the third-order approximation is not distinguishable to    the numerical one. In the low panel  of Fig.\ref{fig1}, we also display the relative error ($\%$) of the three approximate solutions, in comparing them with the exact solution. 

\begin{figure}[t]
\centering
	{\includegraphics[width=75mm]{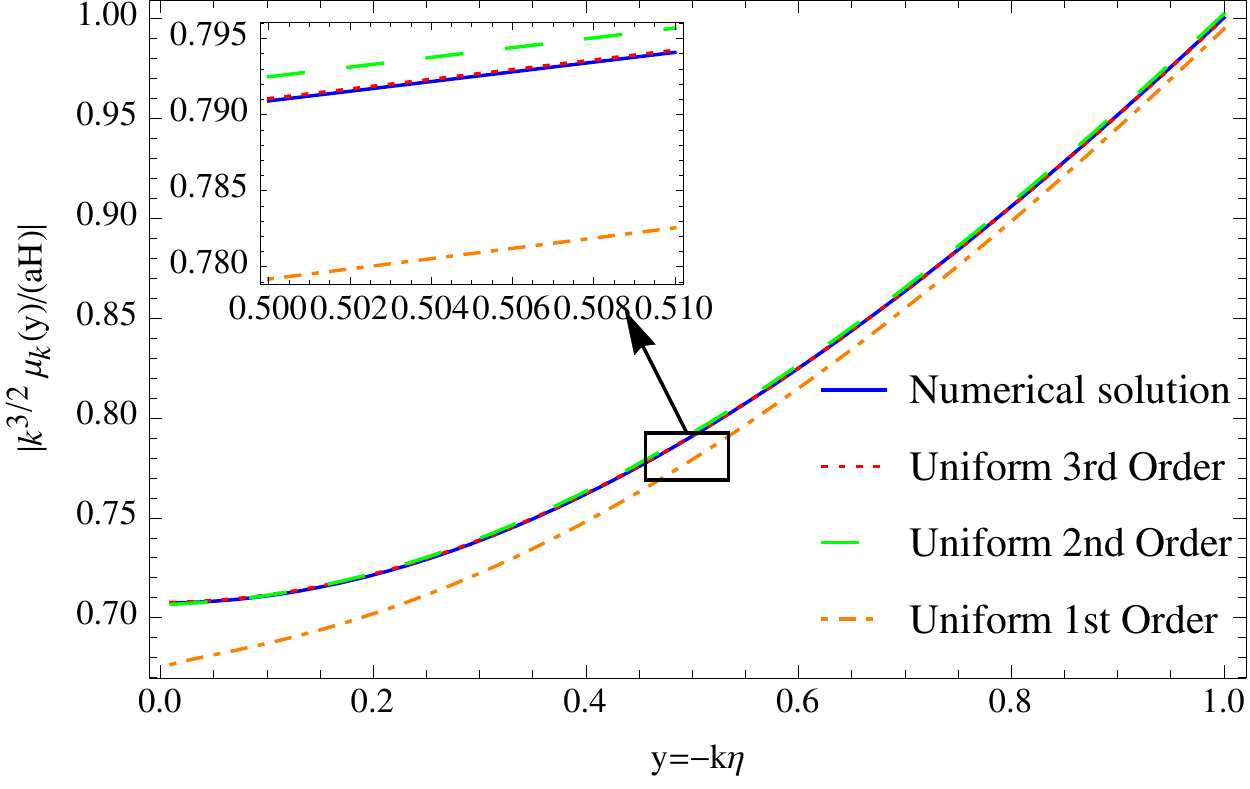}}
	{\includegraphics[width=75mm]{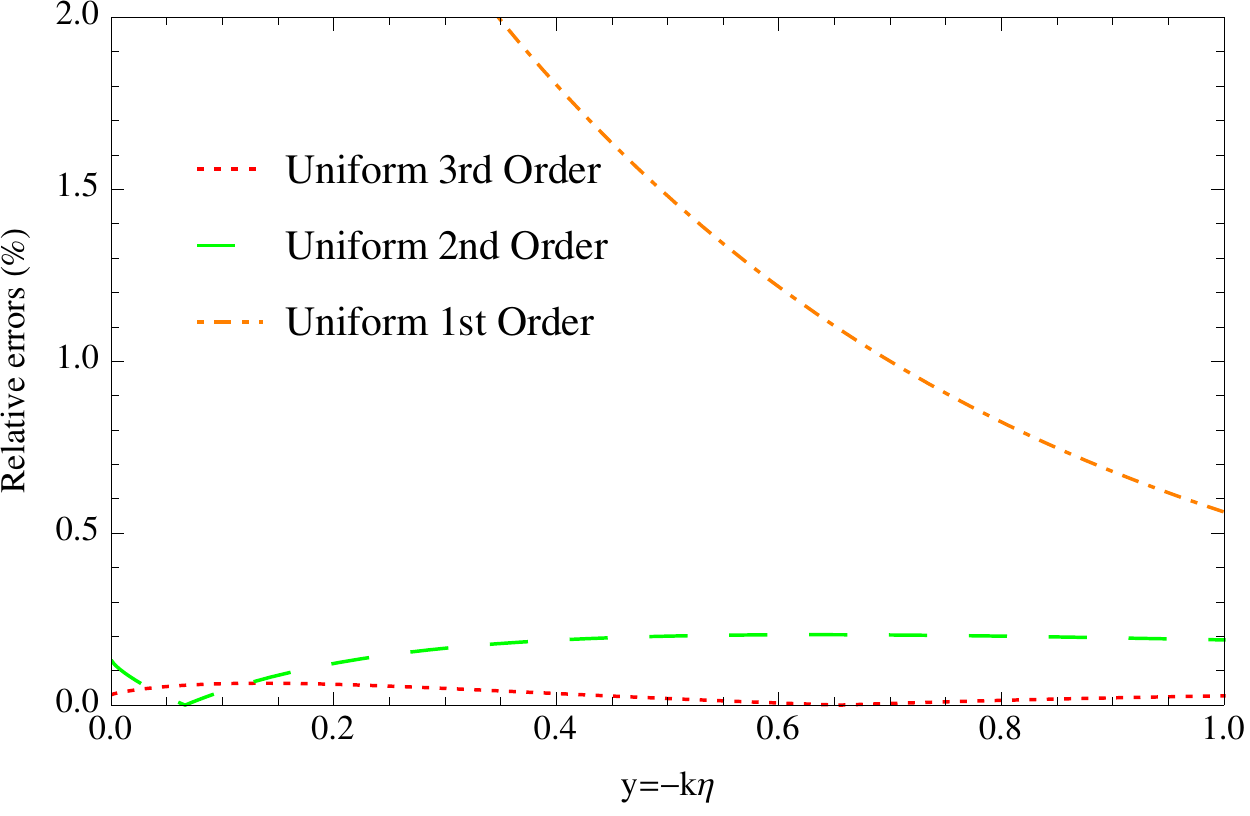}}
	\caption{The mode function $|k^{3/2}\mu_k(y)/(aH)|$ and the corresponding relative errors of the first- second- and third-order approximations of the uniform asymptotic method presented in this paper
 as well as that of   the numerical (exact) solution. (a) Top panel: the numerical solution (blue solid curve) and the approximate solutions of the third-order (red dotted curve), second-order (green dashed curve), and first-order approximation (orange dot-dashed curve). (b) Low panel: The relative errors of the approximate solutions of the third-order (red dotted curve), second-order (green dashed curve), and first-order approximations (orange dot-dashed curve). In drawing the above figures, we have set $\nu=3/2$, $\omega_k^2(\eta)=k^2$ and the mode function was initially at Bunch-Davies vacuum.}
\lb{fig1}
\end{figure}

\section{Power spectra and spectral indices up to the third-order }
\renewcommand{\theequation}{3.\arabic{equation}} \setcounter{equation}{0}

With the approximate solution given in the last section, now let us begin to calculate  the power spectra and spectral indices from the approximate solution.
We assume that the universe was initially at the adiabatic vacuum,
\bqn
\lim_{y\to0^+} \frac{1}{\sqrt{2 \omega_k(\eta)}} e^{-i \int \omega_k(\eta) d\eta},
\eqn
and one needs to match this initial state with the approximate solution (\ref{appro}).  However, the approximate solution (\ref{appro}) involves many high-order terms, which are  complicated
and not easy to handle. In order to simplify them, we  first study their behavior in the limit $y\rightarrow +\infty$. Let us  start with  the $B_0(\xi)$ term in Eq.(\ref{AB}), which satisfies
\bqn
B_0(\xi)=-\frac{1}{2\sqrt{-\xi}} \int_{0}^{\xi} \frac{\psi(v)}{\sqrt{-v}}dv=-\frac{\mathscr{H}(\xi)}{2\sqrt{-\xi}},
\eqn
where $\mathscr{H}(\xi)\equiv \int_{0}^{\xi} dv \psi(v)/|v|^{1/2}$ is the associated error control function of the approximate solution (\ref{appro}), and
in the above we have used $A_0(\xi)=1$. The error control function $\mathscr{H}(\xi)$ is well behaved around the turning point $\bar y_0$ and converges  when $y\to +\infty$. As a result, we have
\bqn
\lim_{y\rightarrow +\infty} B_0(\xi) =-\frac{\mathscr{H}(-\infty)}{2\sqrt{-\xi}}.
\eqn
Then, let us  turn to  $A_1$, which is
\bqn
A_1(\xi)=-\frac{1}{2} B_0'(\xi)+\frac{1}{2}\int_{0}^{\xi} \psi(v)B_0(v)dv.
\eqn
In the limit $y\rightarrow +\infty$,  $B_0'(\xi)$ vanishes, and we find
\bqn
\lim_{y\to+\infty} A_1(\xi) &=&-\frac{1}{2} \int_{0}^{\xi} \frac{\psi(v)}{\sqrt{-v}} \left[\frac{1}{2}\int_{0}^{v} \frac{\psi(u)}{\sqrt{-u}} du\right] dv\nb\\
&=&-\frac{1}{2} \left[\frac{\mathscr{H}(-\infty)}{2}\right]^2.
\eqn
Note that in the above we have used the formula
\bqn
&&n! \int_{\xi_0}^{\xi} f(\xi_n) \int_{\xi_0}^{\xi_n} f(\xi_{n-1})\cdots \int_{\xi_0}^{\xi_2}f(\xi_1)d\xi_1d\xi_2\cdots d\xi_n\nb\\
&&~~~~~~~~~~~~~~~~~~~~~~~~~~~~~~=\left[\int_{\xi_0}^{\xi} f(v)dv\right]^n.
\eqn
Thus,  up to the third-order, we have
\bqn
A_0(\xi)+\frac{A_1(\xi)}{\lambda^2} &=&1-\frac{1}{2\lambda^2} \left[\frac{\mathscr{H}(-\infty)}{2}\right]^2+\mathcal{O}\left(\frac{1}{\lambda^3}\right),\;\;\;\;\;\nb\\
\frac{B_0(\xi)}{\lambda}&=& -\frac{1}{\sqrt{-\xi}} \frac{\mathscr{H}(-\infty)}{2\lambda}+\mathcal{O}\left(\frac{1}{\lambda^3}\right).
\eqn

Thus, using the asymptotic form of Airy  functions in the limit $\xi\to -\infty$, and comparing the solution $\mu_k(y)$ with the initial state, we obtain
\bqn
\alpha_0=\sqrt{\frac{\pi}{2k}} \frac{1}{(A_0+A_1/\lambda^2)-i \sqrt{-\xi} B_0/\lambda},\nb\\
\beta_0=i \sqrt{\frac{\pi}{2k}} \frac{1}{(A_0+A_1/\lambda^2)-i \sqrt{-\xi} B_0/\lambda},
\eqn
where we have
\bqn
(A_0+A_1/\lambda^2)-i \sqrt{-\xi} B_0/\lambda =(1+\mathcal{O}(1/\lambda^3)) e^{i\theta},
\eqn
here $\theta$ is an irrelevant phase factor, and without loss of generality, we can set $\theta=0$. Thus, we  finally get
\bqn
\alpha_0=\sqrt{\frac{\pi}{2k}},\;\;\;\beta_0=i\sqrt{\frac{\pi}{2k}}.
\eqn

After determining the coefficients $\alpha_0$ and $\beta_0$, we can calculate the power spectra of the perturbations. As $y \rightarrow 0$, only  the growing mode is relevant, thus we have
\bqn
\mu_k(y)&\simeq& \beta_0 \left(\frac{\xi}{\hat{g}(y)}\right)^{1/4}\Bigg[\text{Bi}(\lambda^{2/3}\xi)\sum_{s=0}^{+\infty} \frac{B_s(\xi)}{\lambda^{2s}}\nb\\
&&\;\;\;\;\;\;\;\;\;\;\;\;+\frac{\lambda^{2/3}\text{Bi}'(\lambda^{2/3}\xi)}{\lambda^2}\sum_{s=0}^{+\infty}\frac{B_s(\xi)}{\lambda^{2s}}\Bigg].\;\;\;\;\;\;\;\;
\eqn
In order to calculate  the power spectra to higher order, let us first consider the $B_0(\xi)$ term, which satisfies
\bqn
\lim_{y\to 0 }B_0(\xi)=\frac{1}{2\xi^{1/2}} \int_{0}^{\xi} \frac{\psi(v)}{v^{1/2}}dv=\frac{\mathscr{H}(+\infty)}{2\xi^{1/2}}.
\eqn
In the above we had used the relation $\xi^{1/2}d\xi=-\sqrt{\hat{g}}dy$. Knowing the $B_0$ term, we can get the $A_1$ term, which is
\bqn
\lim_{y\to 0} A_1(\xi)&=&\frac{1}{4} \int_{0}^{\xi} \frac{\psi(v)}{v^{1/2}}  \int_{0}^{v} \frac{\psi(u)}{u^{1/2}} du dv\nb\\
&=&\frac{1}{2} \left[\frac{\mathscr{H}(+\infty)}{2}\right]^2.
\eqn

Thus up to the third order and considering the asymptotic forms of the Airy functions in the limit $\xi\rightarrow +\infty$, we find
\bqn
\lim_{y\rightarrow 0} \mu_k(y)&=& \frac{\beta_0 e^{\frac{2}{3}\lambda \xi^{2/3} } }{ \lambda^{1/6} \hat{g}^{1/4} \pi^{1/2} }
\Bigg[1+\frac{\mathscr{H}(+\infty)}{2\lambda}+\frac{\mathscr{H}(+\infty)^2}{8\lambda^2}\nb\\
&&~~~~~~~~~~~~~~~~~~~~~~~~+\mathcal{O}(1/\lambda^3)\Bigg].
\eqn
Then,  the power spectra can be calculated, and is given by
\bqn\lb{pw}
\Delta^2(k)&\equiv& \frac{k^3}{2\pi^2} \left|\frac{\mu_k(y)}{z}\right|^2_{y\to 0^{+}}\nb\\
&\simeq&\frac{k^2}{4\pi^2}\frac{-k\eta}{z^2(\eta) \nu(\eta)}\exp\left(2\int_y^{\bar y_0}\sqrt{\hat{g}(\hat{y})}d\hat{y}\right)\nb\\
&&\;\times \left[1+\frac{\mathscr{H}(+\infty)}{\lambda}+\frac{\mathscr{H}^2(+\infty)}{2\lambda^2}+\mathcal{O}(1/\lambda^3)\right].\nb\\
\eqn

It should be noted that the general expressions of the power spectra have been obtained in \cite{uniformPRL}  up to the second-order,
while in the above expressions the last term in the square brackets represents the third-order approximation.

From the power spectra presented above, one can get the general expression of the spectral indices, which now is given by
\bqn\lb{spectral indices}
n-1&\equiv&\frac{d\ln \Delta^2(k)}{d\ln k}\nb\\
&\simeq&3+2 \int_y^{\bar y_0}\frac{d\hat y}{\sqrt{\hat g(\hat y)}}+\frac{1}{\lambda}\frac{d\mathscr{H}(+\infty)}{d\ln k}\nb\\
&&+\mathcal{O}\left(\frac{1}{\lambda^3}\right),
\eqn
and the last term in the above expression represents the second- and third-order approximations.

It should be noted that the above results represent the most general expressions of the power spectra and  spectral indices of perturbations for the case that has only one-turning point.
In the following we shall apply these expressions to some particular backgrounds and dispersion relations, and calculate the power spectra and
spectral indices.

\section{Power spectra and spectral indices with nonlinear power-law dispersion relation in the slow roll inflation }
\renewcommand{\theequation}{4.\arabic{equation}} \setcounter{equation}{0}

In this section, as an application of the general results obtained  in the last section, let us consider a  specific case where the conventional linear dispersion relation is replaced by the nonlinear one given  in Eq.(\ref{disp}). For simplification, in this section and hereafter we set $\lambda=1$ and thus we have $\hat{g}(y)=g(y)$. In order to get a healthy ultraviolet limit one requires $\hat{b}_2>0$. It is convenient to write the nonlinear dispersion relation as
\bqn\lb{omegastar}
\omega_k^2(\eta)=k^2 \left(1-b_1\epsilon_*^2 y^2+b_2 \epsilon_*^4 y^4\right),
\eqn
where $\epsilon_*\equiv H/M_*$ with $H$ representing the Hubble parameter, which is slowly varying during inflation, and $b_1\equiv \hat{b}_1/(a\eta H)^2$, $b_2\equiv \hat{b}_2/(a\eta H)^4$.
Then, it is easy to find that
\bqn\lb{gofy}
g(y) =\frac{\nu^2}{y^2}-1+b_1\epsilon_*^2 y^2-b_2 \epsilon_*^4 y^4,
\eqn
where we had used the relation
\bqn
\frac{z''}{z}\equiv \frac{\nu^2(\eta)-1/4}{\eta^2}.
\eqn
Now, let us  determine explicitly the power spectra from Eq.(\ref{pw}). We first need to consider the integral of $\sqrt{g(y)}$. However, with the form (\ref{gofy}), the explicit form of the integral usually cannot be worked out explicitly. Thus,   we adopt the expansion given  in \cite{Zhu2},
\bqn
\sqrt{g(y)} &\simeq& \sqrt{\frac{y_0^2}{y^2}-1} \Bigg\{1-\frac{b_2}{2} (y^2+y_0^2) \epsilon_*^2\nb\\
&&-\left[\frac{b_1^2}{8} (y^2+y_0^2)^2-\frac{b_2}{2} (y^4+y^2 y_0^2+y_0^4)\right] \epsilon_*^4\nb\\
&&\;\;+\mathcal{O}(\epsilon_*^4)\Bigg\}.
\eqn
Note that to derive  the above we had assumed that $\epsilon_*$ is  small.  It should be noted that unlike in \cite{Zhu2} where we had treated all the parameters as constant in the first-order slow roll approximation, here in order to go beyond the first-order, we have to treat all the parameters $y_0,\; \epsilon_*, \;b_1,\;\text{and}\;b_2$ as time-dependent.

Even with the above expansion, the integral still cannot be done explicitly, if the explicit form of $y_0(\eta),\;b_1(\eta),\;b_2(\eta),\;\text{and}\;\epsilon_*(\eta)$ are not known. As discussed  in \cite{uniformPRL,uniformPRD}, the integrand in (\ref{pw}) has a square-root singularity at the turning point, i.e., at the upper integral limit. At the lower limit where $y$ goes to zero,  the integrand vanishes linearly. Thus one expects the main contribution to the integral to arise from the upper limit. With this in mind, one can expand all the slowly varying quantities, for example $y_0(\eta)$,  around the turning point $y(\eta_0)=y_0(\eta_0)$ as
\bqn
\lb{nu expansion one}
y_0(\eta)\simeq \bar{y}_0+\frac{dy_0}{d\eta}\Big|_{\eta_0}(\eta-\eta_0)+\frac{1}{2}\frac{d^2y_0}{d\eta^2} \Big|_{\eta_0}(\eta-\eta_0)^2,\nb\\
\eqn
where $\bar{y}_0=-k \eta_0$ represents the turning point, and in the above we only expanded $y_0(\eta)$ to the second-order. With this kind of expansions we can specify all of the quantities $y_0(\eta),\;b_1(\eta),\;b_2(\eta),\;\epsilon_*(\eta)$,  and then calculate the power spectra, spectral indices, and runnings of the indices to the desired accuracy. However, if one considers the slow roll inflation, as pointed out in \cite{uniformPRD}, the above expansions and the slow roll expansion are not independent. On the other hand, by using the slow roll expansion given in Appendix A, one can see that the higher derivative terms which have been ignored in (\ref{nu expansion one}) also contain second-order terms in the slow roll approximation. This means that the final results shall loose some accuracy at the second-order slow roll approximation. Here we should point out that the above analysis is only valid if one makes use of the slow roll approximation.

An alternative expansion of the above quantities was used in \cite{Martin_K1,Martin_K2}, in which $y_0(\eta)$ is given by
\bqn\lb{exy}
y_0(\eta)&\simeq& \bar y_0+\frac{dy_0}{d\ln(-\eta)}\Big|_{\eta_0} \ln\left(\frac{y}{\bar{y}_0}\right)\nb\\
&&+\frac{1}{2}\frac{d^2y_0}{d\ln^2(-\eta)}\Big|_{\eta_0} \ln^2\left(\frac{y}{\bar{y}_0}\right).
\eqn
From the slow roll expansion of high-order derivatives in $y_0(\eta)$, one can see that
\bqn\lb{nu expansion 2}
\frac{d^ny_0}{d\ln^n (-\eta)} \sim \mathcal{O}(\epsilon^{n+1}) +\mathcal{O}(\epsilon^n)\mathcal{O}(\epsilon_*^2),
\eqn
here $\epsilon$ represents the first-order slow-roll quantities. Thus,  if one only considers the power spectra up to the second-order, the higher terms beyond the second-order  in terms of the slow roll parameters can be safely ignored. With this kind  of expansions, we can obtain the explicit expressions of $\sqrt{g(y)}$, the integral of $\sqrt{g(y)}$, and also the error control function $\mathscr{H}(+\infty)$. We present these results in appendices A and B. In Appendices C and D, we present all the slow roll expansions of $\nu(\eta)$, $y_0(\eta)$, $H(\eta)$, $b_1(\eta)$, etc. In the following we shall use these results to calculate  the power spectra and spectral indices of both scalar and tensor perturbations.

First, let us  consider the scalar perturbations, for which we have
\bqn
z_s(\eta) =\sqrt{2\epsilon(\eta)} a(\eta).
\eqn
Then,  we expand the scalar spectrum (\ref{pw}) in terms of slow-roll parameters ($\epsilon,\;\delta_1,\;\delta_2,\;\delta_3 $, etc, which are all defined in appendix C) and $\epsilon_*$, and   find
\begin{widetext}
\bqn
\Delta_s^2(k) &\simeq& %
\frac{181 \bar{H}^2}{72 e^3 \pi ^2\bar \epsilon} %
\Bigg\{  %
1+\frac{909}{724} \hat b_1 \bar \epsilon _*^2-\frac{81 \left(763 \hat b_1^2-1752 \hat b_2\right) \bar \epsilon _*^4}{28960} \nb\\%
&&+\left[4 \ln2-\frac{630}{181}+ \left(\frac{2727}{362}\ln2-\frac{5769 }{724}\right)\hat b_1 \bar \epsilon _*^2-\frac{27  \left(2 \hat b_1^2 (8348+11445 \ln2)+\hat b_2 (14591-52560 \ln2)\right)}{36200}\bar \epsilon _*^4\right]\bar \epsilon \nb\\ %
&&+\left[\ln4-\frac{134}{181}+\frac{3(157+606 \ln2)\hat b_1}{724}  \bar \epsilon _*^2-\frac{27  \left(\hat b_1^2 (1909+4578 \ln2)-2 \hat b_2 (3163+5256 \ln2)\right)}{28960}\bar \epsilon _*^4\right]\bar \delta_1\nb\\ %
&&+\Bigg[\frac{\pi ^2}{3}-\frac{6367}{1629}+4 \ln^22-\frac{536 \ln2}{181}\Bigg] \bar \epsilon^2
+\Bigg[\frac{5 \pi ^2}{12}-\frac{9374}{1629}+3 \ln^22-\frac{40 \ln2}{181}\Bigg]\bar \epsilon \bar \delta_1\nb\\
&&+\Bigg[-\frac{\pi ^2}{12}-\frac{26}{1629}+3 \ln^22-\frac{402 \ln2}{181}\Bigg] \bar \delta_1^2+\Bigg[\frac{543 \pi ^2-688-1629 \ln^24+4824 \ln2}{6516}\Bigg]\bar \delta_2
\Bigg\}.
\eqn
The corresponding scalar spectral index can be calculated from Eq.(\ref{spectral indices}), which is given by
\bqn
n_s&=&1-4 \bar \epsilon-2 \bar \delta _1-\frac{5\hat b_1}{2} \bar \epsilon _*^2 \bar \epsilon+\frac{9}{10} \left(13 \hat b_1^2-22 \hat b_2\right) \bar \epsilon  \bar \epsilon _*^4\nb\\
&&+\Bigg[8 \ln2-\frac{296}{27}-\frac{23}{2} \bar \epsilon _*^2 \hat b_1+\frac{3}{200}\bar \epsilon _*^4 \left(-1607 \hat b_1^2+1560 \ln2 \hat b_1^2+3068 \hat b_2+1410 \ln2 \hat b_2\right)\Bigg]\bar\epsilon^2\nb\\
&&+\Bigg[10\ln2-\frac{262}{27}-\frac{3 \bar\epsilon _*^4}{800}  \left(11977 \hat b_1^2+6240 \ln2 \hat b_1^2-10918 \hat b_2-26760 \ln2 \hat b_2\right)+\bar\epsilon _*^2 \left(5 \ln2 \hat b_2-\frac{53 \hat b_1}{4}\right)\Bigg]\bar\epsilon\bar\delta_1\nb\\
&&+\Bigg[\left(\frac{20}{27}-\ln4\right)\Bigg]\bar\delta_1^2+\Bigg[\bar\epsilon _*^4 \left(\frac{153 \hat b_2}{16}-\frac{171 \hat b_1^2}{32}\right)+\frac{19}{12} \epsilon _*^2 \hat b_1+\left(\ln4-\frac{20}{27}\right)\Bigg]\bar\delta_2\nb\\
&&+\Bigg[\frac{4 \pi ^2}{3}-\frac{1204}{27}-8 \ln2 \ln4+\frac{700 \ln^24}{27}\Bigg]\bar\epsilon^3+\Bigg[\frac{31 \pi ^2}{12}-\frac{3293}{54}-31 \ln^22+\frac{2240 \ln2}{27}\Bigg]\bar\epsilon^2\bar\delta_1\nb\\
&&+\Bigg[\frac{\pi ^2}{4}-\frac{15}{2}-3 \ln^22+\frac{56 \ln2}{9}\Bigg]\bar\epsilon\bar\delta^2+\Bigg[\frac{7 \pi ^2}{12}-\frac{91}{18}-7 \ln^22+\frac{356 \ln2}{27}\Bigg]\bar\epsilon\bar\delta_2\nb\\
&&+\Bigg[\frac{\pi ^2}{6}-\frac{25}{81}-2 \ln^22+\frac{40 \ln2}{27}\Bigg]\bar\delta_1^3+\Bigg[-\frac{\pi ^2}{4}+\frac{31}{162}+3 \ln^22-\frac{20 \ln2}{9}\Bigg]\bar\delta_1 \bar\delta_2\nb\\
&&+\Bigg[\frac{\pi ^2}{12}+\frac{19}{162}-\frac{1}{4} \ln^24+\frac{20 \ln2}{27}\Bigg]\bar\delta_3.
\eqn
Similarly,  for the tensor perturbations we have $z_t(\eta)=a(\eta)$,  and
\bqn
\Delta_t^2(k) &\simeq & \frac{181 \bar H^2}{36 e^3 \pi ^2} \Bigg\{\left(1+\frac{909}{724} \hat b_1 \bar \epsilon _*^2-\frac{81 \left(763 \hat b_1^2-1752 \hat b_2\right) \bar \epsilon _*^4}{28960}\right)\nb\\
&&+\Bigg[\ln4-\frac{496}{181}+ \left(\frac{909}{181}  \ln2-\frac{1560 }{181}\right)\hat b_1\bar \epsilon _*^2+\left(\frac{1545453 \hat b_1^2}{144800}+\frac{1214919 \hat b_2}{72400}+\frac{185409 \hat b_1^2 \ln2}{14480}-\frac{53217 \hat b_2 \ln2}{1810}\right)\bar \epsilon_*^4\Bigg]\bar \epsilon\nb\\
&&+\left(\frac{\pi ^2}{6}-\frac{6635}{1629}+\ln4\right)\bar \epsilon^2
+\left(\frac{\pi ^2}{6}-\frac{9272}{1629}-2 \ln^22+\frac{992 \ln2}{181}\right)\bar \delta_1 \bar \epsilon\Bigg\}.
\eqn
The corresponding tensor spectral index reads
\bqn
n_t&=&\frac{117}{10} \bar\epsilon \bar \epsilon _*^4 \hat b_1^2-\frac{5}{2} \bar\epsilon \bar \epsilon _*^2 \hat b_1-\frac{99}{5} \bar\epsilon \bar \epsilon _*^4 \hat b_2-2 \bar\epsilon\nb\\
&&+\Bigg[\frac{3}{400} \bar\epsilon _*^4 \left((3120 \ln2-589) \hat b_1^2+2 (2993+1410 \ln2) \hat b_2\right)-7 \bar\epsilon _*^2 \hat b_1-\frac{202}{27}+4\ln2\Bigg]\bar\epsilon^2\nb\\
&&+\Bigg[-\frac{1}{400} 3 \bar\epsilon _*^4 \left((2651+3120 \ln2) \hat b_1^2-2 (2017+6690 \ln2) \hat b_2\right)+\frac{1}{3} \bar \epsilon _*^2 (15\ln2-31) \hat b_1-\frac{148}{27}+4\ln2\Bigg]\bar\epsilon\bar\delta_1 \nb\\
&&+\Bigg[\frac{2 \pi ^2}{3}-\frac{3106}{81}-8 \ln^22+\frac{916 \ln2}{27}\Bigg]\bar\epsilon^3+\Bigg[\frac{7 \pi ^2}{6}-\frac{4219}{81}-14 \ln^22+\frac{1360 \ln2}{27}\Bigg]\bar\epsilon^2\bar\delta_1\nb\\
&&+\Bigg[\frac{\pi ^2}{6}-\frac{425}{81}-2 \ln^22+\frac{148 \ln2}{27}\Bigg]\bar\epsilon\bar\delta_1^2+\Bigg[\frac{\pi ^2}{6}-\frac{425}{81}-2 \ln^22+\frac{148 \ln2}{27}\Bigg]\bar\epsilon\bar\delta_2.
\eqn
It should be noted that all the above expressions are evaluated at the turning point $y=\bar y_0$.

\section{Power spectra, spectral indices, and running of indices with the linear dispersion relation in the slow roll inflation }
\renewcommand{\theequation}{5.\arabic{equation}} \setcounter{equation}{0}

The results given  in the last section can be easily reduced to the case for a linear dispersion relation $\omega_k(\eta)=k^2$ by setting $\hat b_1=0=\hat b_2$. In the following we present the expressions of the power spectra, spectral indices, and  runnings of the indices for both scalar and tensor perturbations.

\subsection{Scalar Perturbations}

For the scalar power spectrum, we get
\bqn
\Delta_s^2(k) &\simeq& %
\frac{181 \bar{H}^2}{72 e^3 \pi ^2\bar \epsilon} %
\Bigg\{ 1+\left(4 \ln2-\frac{630}{181}\right)\bar \epsilon+\left(\ln4-\frac{134}{181}\right)\bar \delta_1+\left(\frac{\pi ^2}{3}-\frac{6367}{1629}+4 \ln^22-\frac{536 \ln2}{181}\right)\bar \epsilon^2\nb\\
&&~~~~~~~~~~~~~~~~~~+\left(\frac{5 \pi ^2}{12}-\frac{9374}{1629}+3 \ln^22-\frac{40 \ln2}{181}\right) \bar \epsilon \bar \delta_1+\left(-\frac{\pi ^2}{12}-\frac{26}{1629}+3 \ln^22-\frac{402 \ln2}{181}\right) \bar \delta_1^2\nb\\
&&~~~~~~~~~~~~~~~~~~+
\left(\frac{\pi ^2}{12}-\frac{172}{1629}-\ln^22+\frac{134 \ln2}{181}\right) \bar \delta_2
\Bigg\}.
\eqn

The spectral index $n_{s}$ can be expressed as
\bqn
n_s-1&=&
-4\bar \epsilon
-2 \bar \delta _1
+\left(8 \ln2-\frac{296}{27}\right)\bar\epsilon^2
+\left(\ln(1024)-\frac{262}{27}\right)\bar\epsilon\bar\delta_1
+\left(\frac{20}{27}-\ln4\right)\bar\delta_1^2
+\left(\ln4-\frac{20}{27}\right)\bar\delta_2
\nb\\&&
+\left(\frac{4 \pi ^2}{3}-\frac{1204}{27}-4 \ln^24+\frac{700 \ln4}{27}\right)\bar\epsilon^3
+\left(\frac{31 \pi ^2}{12}-\frac{3293}{54}-31 \ln^22+\frac{2240 \ln2}{27}\right)\bar\epsilon^2\bar\delta_1
\nb\\&&
+\left(\frac{\pi ^2}{4}-\frac{15}{2}-3 \ln^22+\frac{56 \ln2}{9}\right)\bar\epsilon\bar\delta^2
+\left(\frac{7 \pi ^2}{12}-\frac{91}{18}-7 \ln^22+\frac{356 \ln2}{27}\right)\bar\epsilon\bar\delta_2
\nb\\&&
+\left(\frac{\pi ^2}{6}-\frac{25}{81}-2 \ln^22+\frac{40 \ln2}{27}\right)\bar\delta_1^3
+\left(-\frac{\pi ^2}{4}+\frac{31}{162}+3 \ln^22-\frac{20 \ln2}{9}\right)\bar\delta_1 \bar\delta_2
\nb\\&&
+\left(\frac{\pi ^2}{12}+\frac{19}{162}-\frac{1}{4} \ln^24+\frac{20 \ln2}{27}\right)\bar\delta_3.
\eqn

The running of the spectral index is given by
\bqn
\alpha_s&\simeq &
-8 \bar \epsilon ^2
-10 \bar \delta _1 \bar \epsilon
+2 \bar \delta_1^2
-2\bar \delta_2
+ \left(32 \ln2-\frac{1400}{27}\right)\bar \epsilon ^3
+ \left(62 \ln2-\frac{2240}{27}\right)\bar \delta _1 \bar \epsilon ^2
+ \left(14 \ln2-\frac{356}{27}\right)\bar \delta _2\bar  \epsilon
\nb\\&&
+ \left(6 \ln2-\frac{56}{9}\right)\bar \delta _1^2 \bar \epsilon
+ \left(4 \ln2-\frac{40}{27}\right)\bar \delta _1^3
+ \left(\frac{20}{9}-6 \ln2\right)\bar \delta _1 \bar \delta _2
+\left(2 \ln2-\frac{20}{27}\right)\bar \delta _3
\nb\\&&
+ \left(\frac{\pi ^2}{12}-\frac{19}{162}-\ln^22+\frac{20 \ln2}{27}\right)\bar \delta _4
+ \left(8 \pi ^2-\frac{8480}{27}-96\ln^22+\frac{3088 \ln2}{9}\right)\bar \epsilon ^4
\nb\\&&
+ \left(\frac{251 \pi ^2}{12}-\frac{34457}{54}-251\ln^22+\frac{21274 \ln2}{27}\right)\bar \delta _1 \bar \epsilon ^3
+ \left(\frac{35 \pi ^2}{4}-\frac{425}{2}-105\ln^22+\frac{2518 \ln2}{9}\right)\bar \delta _1^2 \bar \epsilon ^2
\nb\\&&
+ \left(\frac{59 \pi ^2}{12}-\frac{1579}{18}-59\ln^22+\frac{4042 \ln2}{27}\right)\bar \delta _2 \bar \epsilon ^2
+ \left(\frac{\pi ^2}{2}-\frac{317}{27}-6\ln^22+\frac{76 \ln2}{9}\right)\bar \delta _1^3\bar \epsilon
\nb\\&&
+ \left(\frac{\pi ^2}{3}-\frac{214}{27}-4\ln^22+\frac{175\ln4}{27}\right)\bar \delta _1 \bar \delta _2 \bar \epsilon
+ \left(\frac{5 \pi ^2}{6}-\frac{49}{9}-10\ln^22+\frac{470 \ln2}{27}\right)\bar \delta _3 \bar \epsilon
\nb\\&&
+\left(-\frac{\pi ^2}{2}+\frac{61}{27}+6\ln^22-\frac{40 \ln2}{9}\right)\bar \delta _1^4 
+ \left(\pi ^2-\frac{322}{81}-12\ln^22+\frac{80 \ln2}{9}\right)\bar \delta _1^2 \bar \delta _2
\nb\\&&
+ \left(-\frac{\pi ^2}{3}+\frac{2}{27}+4\ln^22-\frac{80 \ln2}{27}\right)\bar\delta _1 \bar \delta _3
+\left(-\frac{\pi ^2}{4}+\frac{247}{162}+3\ln^22-\frac{20 \ln2}{9}\right)\bar \delta _2^2 .
\eqn
Note that the above results are all evaluated at the turning point $\bar y_0$. One can translate them into the expressions evaluated at other points, for example, at the horizon crossing $\eta_\star $ with $a(\eta_\star)H(\eta_\star)=k$. In the subsection D, we present all the results evaluated at horizon crossing and compare them with the results obtained by other methods.

\subsection{Tensor Perturbations}

For the tensor power spectrum, we find
\bqn
\Delta_t^2(k) &\simeq & \frac{181 \bar H^2}{36 e^3 \pi ^2} \Bigg\{1+  \left(\ln4-\frac{496}{181}\right)\bar \epsilon+\left(\frac{\pi ^2}{6}-\frac{6635}{1629}+2 \ln2\right)\bar \epsilon^2+\left(\frac{\pi ^2}{6}-\frac{9272}{1629}-2 \ln^22+\frac{992 \ln2}{181}\right) \bar \delta_1\bar \epsilon\Bigg\}.\nb\\
\eqn
Then, the tensor spectral index $n_{T}$ can be expressed as
\bqn
n_t&=&-2 \bar \epsilon+\left(4\ln2-\frac{202}{27}\right)\bar \epsilon^2+\left(4\ln2-\frac{148}{27}\right)\bar \epsilon\bar \delta_1 +\left(\frac{2 \pi ^2}{3}-\frac{3106}{81}-8 \ln^22+\frac{916 \ln2}{27}\right)\bar\epsilon^3\nb\\
&&+\left(\frac{7 \pi ^2}{6}-\frac{4219}{81}-14 \ln^22+\frac{1360 \ln2}{27}\right)\bar\epsilon^2\bar \delta_1+\left(\frac{\pi ^2}{6}-\frac{425}{81}-2 \ln^22+\frac{148 \ln2}{27}\right)\bar\epsilon\bar\delta_1^2\nb\\
&&+\left(\frac{\pi ^2}{6}-\frac{425}{81}-2 \ln^22+\frac{148 \ln2}{27}\right)\bar\epsilon\bar\delta_2,
\eqn
while the corresponding running  is given by
\bqn
\alpha_t &\simeq &
-4 \bar \delta _1 \bar \epsilon 
-4\bar \epsilon ^2
+\left(16 \ln2-\frac{916}{27}\right)\bar \epsilon^3
+\left(28 \ln2-\frac{1360}{27}\right)\bar \epsilon^2 \bar\delta_1
+\left(4 \ln2-\frac{148}{27}\right)\bar\epsilon \bar \delta_2 
\nb\\&&
+\left(4 \pi ^2-\frac{800}{3}-48 \ln^22+\frac{988\ln4}{9}\right)\bar \epsilon^4
+ \left(\frac{59 \pi ^2}{6}-\frac{44243}{81}-118 \ln^22+\frac{13052 \ln2}{27}\right)\bar\delta _1 \bar \epsilon ^3
\nb\\&&
+ \left(\frac{25 \pi ^2}{6}-\frac{15017}{81}-50 \ln^22+\frac{4780 \ln2}{27}\right)\bar\delta _1^2 \bar \epsilon ^2
+ \left(\frac{11 \pi ^2}{6}-\frac{707}{9}-22 \ln^22+\frac{2060 \ln2}{27}\right)\bar \delta _2 \bar \epsilon ^2
\nb\\&&
+ \left(\frac{\pi ^2}{2}-\frac{425}{27}-6 \ln^22+\frac{148 \ln2}{9}\right)\bar \delta _1 \bar\delta _2\bar \epsilon
+ \left(\frac{\pi ^2}{6}-\frac{425}{81}-2 \ln^22+\frac{148 \ln2}{27}\right)\bar\delta _3 \bar\epsilon.
\eqn

\subsection{Comparing with results obtained from the  Green function method}

In the last subsection, we have obtained the expressions for power spectra, spectral indices, and running of spectral indices for both scalar and tensor perturbations. It should be noted that all these expressions were evaluated at the turning point $\bar y_0$, i.e., $\eta=\eta_0$. However, in the usual treatments,   all expressions were expanded  at the horizon crossing $a(\eta_\star) H(\eta_\star)=k$. In order to compare our results  with the ones obtained by other methods \cite{green1,green2,WKB}, we need to rewrite our expressions in terms of  the quantities evaluated at horizon crossing $\eta_\star$. This can be achieved by using the following expansion
\bqn
f(\eta_0)&\simeq& f(\eta_\star)+\frac{df(\eta)}{d\ln (-\eta)}\Big|_{\eta_\star} \ln{\left(\frac{\eta_0}{\eta_\star}\right)}+\frac{1}{2}\frac{d^2f(\eta)}{d\ln^2(-\eta)}\Big|_{\eta_\star} \ln^2{\left(\frac{\eta_0}{\eta_\star}\right)}+\frac{1}{6} \frac{d^3f(\eta)}{d\ln^3(-\eta)}\Big|_{\eta_\star} \ln^3{\left(\frac{\eta_0}{\eta_\star}\right)},
\eqn
where for scalar perturbations, one can expand $\ln{\left(\frac{\eta_0}{\eta_\star}\right)}$ as
\bqn
\ln\left(\frac{\eta_0}{\eta_\star}\right)
&=&
\ln\left[\frac{\nu_s(\eta_0)}{a(\eta_\star) H(\eta_\star) \eta_\star}\right]\nb\\
&\simeq& 
\ln\frac{3}{2}
+\frac{2}{3}\delta_{\star1}
+\frac{1}{3}\epsilon_\star
+\left(\frac{1}{6}-\frac{8}{3} \ln\frac{3}{2}\right)\epsilon_\star ^2
+\left(\frac{2}{9}-\frac{10}{3} \ln \frac{3}{2}\right)\delta_{\star1} \epsilon_\star
+\left(\frac{2}{3} \ln\frac{3}{2}-\frac{4}{9}\right)\delta_{\star1}^2
+\left(\frac{2}{9}-\frac{2}{3} \ln\frac{3}{2}\right)\delta _{\star2}
\nb\\&&
+\left(-\frac{155}{81}+\frac{16}{3} \ln^2\frac{3}{2}-\frac{40}{3} \ln\frac{3}{2}\right)\epsilon_\star^3
+\left(-\frac{134}{27}+\frac{31}{3} \ln^2\frac{3}{2}-\frac{62}{3} \ln\frac{3}{2}\right)\delta_{\star1} \epsilon_\star^2
+ \left(-\frac{68}{27}+\ln^2\frac{3}{2}-\frac{2}{3} \ln\frac{3}{2}\right)\delta_{\star1}^2 \epsilon_\star
\nb\\&&
+\left(\frac{68}{81}+\frac{2}{3} \ln^2\frac{3}{2}-\frac{8}{9} \ln\frac{3}{2}\right)\delta_{\star1}^3
+\left(-\frac{16}{27}+\frac{7}{3} \ln^2\frac{3}{2}-\frac{10}{3} \ln\frac{3}{2}\right)\delta_{\star2} \epsilon_\star 
+\left(-\frac{20}{27}-\ln^2\frac{3}{2}+\frac{10}{9} \ln\frac{3}{2}\right)\delta_{\star1} \delta_{\star2}
\nb\\&&
+ \left(\frac{1}{3} \ln^2\frac{3}{2}-\frac{2}{9} \ln\frac{3}{2}\right)\delta_{\star3},
\eqn
and for tensor perturbations, we find
\bqn
\ln\left(\frac{\eta_0}{\eta_\star}\right)
&=&\ln\left[\frac{\nu_t(\eta_0)}{a(\eta_\star) H(\eta_\star) \eta_\star}\right]\nb\\
&\simeq& 
\ln\frac{3}{2}
-\frac{\epsilon_\star}{3}
-\left(\frac{5}{18}+\frac{4}{3}\ln\frac{3}{2}\right)\epsilon_\star ^2
-\left(\frac{2}{9}+\frac{4}{3} \ln\frac{3}{2}\right)\delta _{\star1} \epsilon_\star
+\left(-\frac{2}{9}-\frac{16}{9} \ln\frac{3}{2}+\frac{2}{3} \ln^2\frac{3}{2}\right)\epsilon_\star \delta_{\star2}
\nb\\&&
+\left(-\frac{49}{81}+\frac{8}{3} \ln^2\frac{3}{2}-\frac{92}{9} \ln\frac{3}{2}\right)\epsilon_\star ^3 
+\left(-\frac{2}{9}-\frac{16}{9} \ln\frac{3}{2}+\frac{2}{3} \ln^2\frac{3}{2}\right)\delta _{\star1} \epsilon_\star ^2.
\eqn
In the following we shall use the above expansions to translate all the results into the expressions evaluated at the horizon crossing.

\subsubsection{Scalar spectrum and spectral index}

By making use of the above expansions, the power spectrum for scalar perturbations can be cast in the form
\bqn
\Delta_s^2(k)&\simeq& \frac{181H_\star^2}{72 e^3\pi^2 \epsilon_\star } \Bigg\{1+\delta _{\star1} \left(\ln9-\frac{134}{181}\right)+\epsilon_\star  \left(2\ln9-\frac{630}{181}\right)+ \left(\frac{\pi ^2}{3}-\frac{4195}{1629}+4 \ln^23-\frac{536 \ln3}{181}\right)\epsilon_\star ^2\nb\\
&&\;\;\;\;\;\;\;+\delta _{\star1} \epsilon_\star  \left(\frac{5 \pi ^2}{12}-\frac{3944}{1629}+3 \ln^23-\frac{40 \ln3}{181}\right)+\delta _{\star1}^2 \left(-\frac{\pi ^2}{12}+\frac{2146}{1629}+3 \ln^23-\frac{402 \ln3}{181}\right)\nb\\
&&\;\;\;\;\;\;+\delta _{\star2} \left(\frac{\pi ^2}{12}-\frac{172}{1629}-\ln^23+\frac{134 \ln3}{181}\right)\Bigg\}.
\eqn
One can compare the above  expression with the one  obtained by using the Green function method  \cite{green1},
\bqn
\Delta_{\text{Green}}^2(k)&\simeq&\frac{H_\star^2}{8\pi^2 \epsilon_\star }\Bigg\{1+(4\alpha_\star-2)\epsilon_\star+2\alpha_\star \delta_1+\left(4\alpha_\star-23+\frac{7\pi^2}{3}\right)\epsilon_\star^2\nb\\
&&+\left(3\alpha_\star+2\alpha_\star-22+\frac{29\pi^2}{12} \right)\epsilon_\star\delta_{\star1}+\left(3\alpha^2_\star-4+\frac{5\pi^2}{12}\right)\delta_{\star1}^2+\left(-\alpha_\star^2+\frac{\pi^2}{12}\right)\delta_{\star2}\Bigg\},
\eqn
where $\alpha_\star=2-2\ln 2-\gamma$ with the Euler constant $\gamma\simeq 0.577216$.
In Table I, we compare the amplitude and the numerical coefficients of the above two expressions  for the scalar spectrum, from which one can see that they are extremely close to each other, and
essentially the same within the errors allowed.

\begin{table}[htdp]
\caption{Comparing with the Green function method: scalar power spectrum}
\begin{center}
\begin{tabular}{|c|c|c|c|c|c|c|c|}
\hline
Methods & Amplitude & $\epsilon_\star$ & $\delta_{\star1}$ & $\epsilon^{\star2}$ &$\delta_{\star1}\epsilon_\star$ & $\delta_{\star1}^2$ & $\delta_{\star2}$\\
\hline
Uniform Approximation &  $\frac{181H_\star^2}{72 e^3\pi^2 \epsilon_\star }$ & 0.913786 & 1.456893 & 2.28912& 5.06928& 1.67574& 0.323269\\
\hline
Green function method \cite{green1}  & $\frac{H_\star^2}{8\pi^2 \epsilon_\star }$ & 0.918549 & 1.459274 & 2.158558 &4.907929 & 1.709446 & 0.290097\\
\hline
Relative difference & $\sim 0.13\%$ & $0.52\%$& $0.16\%$ &  5.7\% &3.2\% & 2.0\%& 10.3\%\\
\hline
\end{tabular}
\end{center}
\label{table1}
\end{table}%

Now we turn to the scalar spectral index $n_s$, which can be rewritten as
\bqn
n_s-1&\simeq&-2 \delta _{\star1}-4 \epsilon_\star+\epsilon_\star^2 \left(8 \ln3-\frac{296}{27}\right)%
+\delta _{\star1} \epsilon_\star  \left(10 \ln3-\frac{262}{27}\right)%
+\delta _{\star1}^2 \left(\frac{20}{27}-\ln9\right)+\delta _{\star2} \left(\ln9-\frac{20}{27}\right)\nb\\
&&+\epsilon_\star^3 \left(\frac{4 \pi ^2}{3}-\frac{1132}{27}-16 \ln^23+\frac{700 \ln9}{27}\right)+\delta _{\star3} \left(\frac{\pi ^2}{12}+\frac{19}{162}-\ln^23+\frac{20 \ln3}{27}\right)\nb\\
&&+\delta _{\star1} \epsilon_\star ^2 \left(\frac{31 \pi ^2}{12}-\frac{2825}{54}-31 \ln^23+\frac{2240 \ln3}{27}\right)\nb\\
&&+\delta _{\star1}^2 \epsilon_\star  \left(\frac{\pi ^2}{4}-\frac{3}{2}-3 \ln^23+\frac{56 \ln3}{9}\right)
+\delta _{\star2} \epsilon_\star  \left(\frac{7 \pi ^2}{12}-\frac{79}{18}-7 \ln^23+\frac{356 \ln3}{27}\right)\nb\\
&&+\delta _{\star1}^3 \left(\frac{\pi ^2}{6}-\frac{133}{81}-2 \ln^23+\frac{40 \ln3}{27}\right)+\delta _{\star1} \delta _{\star2} \left(-\frac{\pi ^2}{4}+\frac{247}{162}+3 \ln^23-\frac{20 \ln3}{9}\right).
\eqn
In Table II, we compare numerical coefficients for the first-order and second order terms in the scalar spectral index, with the results obtained by using the Green function method \cite{green1}. From the table one can see that both results are essentially identically within the errors allowed.

\begin{table}[htdp]
\caption{Compare with Green function method: scalar spectral index}
\begin{center}
\begin{tabular}{|c|c|c|c|c|c|c|c|}
\hline
Methods &  $\epsilon_\star$ & $\delta_{\star1}$ & $\epsilon_\star^2$ &$\delta_{\star1}\epsilon_\star$ & $\delta_{\star1}^2$ & $\delta_{\star2}$\\
\hline
Uniform Approximation  & -2 & -4 & -2.174065& 1.282419& -1.456484& 1.456484\\
\hline
Green function method \cite{green1}   & -2 & -4 & -2.162903 &1.296372 & -1.459274 & 1.459274\\
\hline
Relative difference & $0\%$ & $0\%$& $0.5\%$ &  1.1\% &0.2\% & 0.2\%\\
\hline
\end{tabular}
\end{center}
\label{table2}
\end{table}%

\subsubsection{Tensor spectrum and spectral index}

For the tensor power spectrum, we find
\bqn
\Delta_t^2(k)&\simeq& \frac{181 H_\star^2}{36 e^3 \pi ^2} \Bigg\{1+  \left(\ln9-\frac{496}{181}\right)\epsilon_\star +\left(\frac{\pi ^2}{6}-\frac{7721}{1629}+\ln9\right)\epsilon_\star^2+\left(\frac{\pi ^2}{6}-\frac{9272}{1629}-2 \ln^23+\frac{992 \ln3}{181}\right)\epsilon_\star \delta_{\star1}\Bigg\},\nb\\
\eqn
and the expression from the Green function method \cite{green2} is
\bqn
\Delta_{\text{Green}}^2(k) &\simeq& \frac{H_\star^2}{4\pi^2} \Bigg\{1+(2\alpha-2) \epsilon_\star+\left(2\alpha-9+\frac{2\pi^2}{3}\right)\epsilon_\star^2+\left(-2\alpha^2+4\alpha-4+\frac{\pi^2}{6}\right)\epsilon_\star \delta_{\star1}\Bigg\}.
\eqn
In Table III, we compare the amplitude and the numerical coefficients of the above two expressions for the tensor spectrum,  and find  that the two results are essentially the same.

\begin{table}[htdp]
\caption{Compare with Green function method: tensor power spectrum}
\begin{center}
\begin{tabular}{|c|c|c|c|c|c|c|c|}
\hline
Methods & Amplitude & $\epsilon_\star$ & $\epsilon_\star^2$ &$\delta_{\star1}\epsilon_\star$ \\
\hline
Uniform Approximation &  $\frac{181H_\star^2}{36 e^3\pi^2 }$ & -0.543107 & -0.897559 & -0.439676\\
\hline
Green function method \cite{green2} & $\frac{H_\star^2}{4\pi^2 }$ & -0.540726 & -0.967524 & -0.504813 \\
\hline
Relative difference & $\sim 0.13\%$ & $0.4\%$& $7\%$ &  13\% \\
\hline
\end{tabular}
\end{center}
\label{table3}
\end{table}%

Now we turn to consider the spectral index, which can be written as
\bqn
n_t &\simeq&  -2 \epsilon_\star+\left(2\ln9-\frac{202}{27}\right) \epsilon_\star^2+\left(2\ln9-\frac{148}{27}\right)\epsilon_\star\delta_{\star1}+\left(\frac{2 \pi ^2}{3}-\frac{3214}{81}-8 \ln^23+\frac{916 \ln3}{27}\right)\epsilon_\star^3\nb\\
&&+\left(\frac{7 \pi ^2}{6}-\frac{4327}{81}-14 \ln^23+\frac{1360 \ln3}{27}\right)\epsilon_\star^2\delta_{\star1}+\left(\frac{\pi ^2}{6}-\frac{425}{81}-2 \ln^23+\frac{148 \ln3}{27}\right)\delta_{1\star}^2 \epsilon_{\star}\nb\\
&&+\left(\frac{\pi ^2}{6}-\frac{425}{81}-2 \ln^23+\frac{148 \ln3}{27}\right)\epsilon_\star\delta_{\star2}.
\eqn
In Table IV, we compare the numerical coefficients for the first-order and second-order terms in the tensor spectral index, with the results obtained by using the Green function method \cite{green2}.

\begin{table}[htdp]
\caption{Compare with Green function method: tensor spectral index}
\begin{center}
\begin{tabular}{|c|c|c|c|c|c|c|c|}
\hline
Methods &  $\epsilon_\star$ &  $\epsilon_\star^2$ &$\delta_{\star1}\epsilon_\star$ \\
\hline
Uniform Approximation  & -2  & -3.08703& -1.087032\\
\hline
Green function method  \cite{green2} & -2  & -3.08145 &-1.08145 \\
\hline
Relative difference & $0\%$ & $0.2\%$ &  0.5\% \\
\hline
\end{tabular}
\end{center}
\label{table4}
\end{table}%

With the scalar spectrum and tensor spectrum given in the above, we find that the tensor-to-scalar ratio is expressed as
\bqn
r&\simeq& 16 \epsilon_{\star} 
\Bigg\{
1
+\left(\frac{134}{181}-\ln9\right) \epsilon_\star
+ \left(\frac{134}{181}-\ln9\right)\delta _{\star1}
+ \left(-\frac{\pi ^2}{4}-\frac{4776}{32761}+7 \ln^23-\frac{1300 \ln3}{181}\right)\epsilon_\star \delta_{\star1}
\nb\\&&
~~~~~~~~+ \left(\frac{\pi ^2}{12}-\frac{226822}{294849}+\ln^23-\frac{134 \ln3}{181}\right)\delta _{\star1}^2
+ \left(-\frac{\pi ^2}{12}+\frac{172}{1629}+\ln^23-\frac{134 \ln3}{181}\right)\delta _{\star2}
\nb\\&&
~~~~~~~~+ \left(-\frac{\pi ^2}{6}+\frac{121574}{294849}+4 \ln^23-\frac{898 \ln3}{181}\right)\epsilon_{\star}^2\Bigg\}.
\eqn

Finally, we note that,  although we only compared  our results with those obtained by the Green function method \cite{green1,green2}, our results are also comparable  with results obtained by the WKB approximation method \cite{WKB}. In fact, in Ref. \cite{Martin_K2}, the authors checked various versions of the power spectra, including those obtained by the first-order uniform approximation, Green function method, and also improved-WKB method, and found that they are all essentially the same.

\end{widetext}

\section{Conclusions}
\renewcommand{\theequation}{6.\arabic{equation}} \setcounter{equation}{0}

In this paper, by using the {\em uniform asymptotic approximation method}, we have calculated the power spectra and  spectral indices of both scalar and tensor perturbations. We have implemented the high-order uniform approximations and presented the general expressions of both  power spectra and spectral indices up to the third-order in the uniform approximations. To see the quantum gravitational  effects,  we have studied  the nonlinear power-law dispersion relation and calculated explicitly the power spectra and spectral indices of scalar and tensor perturbations in the slow-roll inflation. Furthermore, we  have also considered the GR limit of the power spectra and spectral indices, and calculate the corresponding  runnings of the spectral indices. From these results one can see that the uniform approximation method presents a powerful approach to calculate the inflationary power spectra and spectral indices.

The precision of our results can be shown by two approaches. First, restricting ourselves to the relativistic case, we have compared  our expressions of power spectra and spectral indices of scalar and tensor perturbations   with those obtained previously by the Green function method. From Tables I-IV, one can see that the results of the two methods are extremely close to each other. In fact, they are the same within the errors allowed. Second,  in Fig.\ref{fig1} we present our analytical solution of the mode function to the first-, second-, and third-order approximations, respectively, and then compare them with the numerical (exact) evolution of the mode function. From there one can see clearly that our analytical solution up to the second-order approximation  is already extremely closed to the numerical one. 
In the general case, the rigorous approach to determine the precision of our results is to analyze the error bounds presented in (\ref{error}). However, as the values of all the parameters usually are not known, it is very difficult to get exact numerical values about the errors. But, as all the relevant parameters, like $\epsilon, \epsilon_*$, etc, are small quantities, it is effective and useful to use the error bounds for $\nu_{s,t}=3/2$ and $\epsilon_*=0$, which have been discussed rigorously in \cite{uniformPRL, uniformPRD}. The real error bounds should be not far from it. In table V, we present the expected errors of the power spectra in our uniform approximations.

The uniform asymptotic approximation method presented in this paper has at least two major advantages over  other methods proposed so far in the literature. First,  the error bounds  are explicitly constructed order by order, so that at each order the errors are well under our control. This is in contrast to all the other methods proposed so far in which the errors are not known.  However, knowing the errors at each order  is essential for the further improvement of the accuracy of the calculations of the mode function, power spectra, spectral indices and runnings.  Second, such an approximation can be easily extended to more interesting cases including cases with milt-turning and/or high-order turning points, which usually arise when the gravitational quantum effects are taken into account, as shown in the Introduction. 

In review of all the above, one can see clearly that   the uniform asymptotic approximation method indeed provides a powerful tool to address various questions in   inflationary models about the gravitational quantum effects  in the framework of string/M theory, loop quantum gravity, Horava-Lifshitz gravity, and so on. It would be extremely important to find some observational signatures of the gravitational quantum effects
for the forthcoming observations. 

Finally, it should be noted that in this paper we have only considered the cases with one single-turning point. It is very interesting to study the quantum effects in the cases with several different turning points or one multiple-turning point \cite{uniformZhu}. Meantime, as our results are very general, it would be very interesting to apply them to other cosmological models.

\begin{table}[htdp]
\caption{Errors to be expected in the uniform approximation}
\begin{center}
\begin{tabular}{|c|c|c|c|c|c|}
\hline
Quantity & $1$st-order &  $2$nd-order  & $3$th-order \\
\hline
Power spectrum: $\Delta^2(k)$ & $\lesssim 10\%$  & $\lesssim 1\%$ & $\lesssim 0.1\%$\\
\hline
\end{tabular}
\end{center}
\label{default}
\end{table}%

\section*{Acknowledgements}

This work is supported in part by DOE, DE-FG02-10ER41692 (AW),
Ci\^encia Sem Fronteiras, No. 004/2013 - DRI/CAPES (AW),
NSFC No. 11375153 (AW), No. 11173021 (AW), No. 11047008 (TZ), No. 11105120 (TZ), and No. 11205133 (TZ).

\section*{Appendix A: Expansion of $\sqrt{g(y)}$}
\renewcommand{\theequation}{A.\arabic{equation}} \setcounter{equation}{0}
In this section, we present the expression of the expansion of $\sqrt{g(y)}$. First, we can expand $\sqrt{g(y)}$ in terms of $\epsilon_*$ as
\bqn\lb{gexpansion}
\sqrt{g(y)} &\simeq& \sqrt{\frac{y_0^2}{y^2}-1} \Bigg\{1-\frac{b_2}{2} (y^2+y_0^2) \epsilon_*^2\nb\\
&&-\left[\frac{b_1^2}{8} (y^2+y_0^2)^2-\frac{b_2}{2} (y^4+y^2 y_0^2+y_0^4)\right] \epsilon_*^4\nb\\
&&\;\;+\mathcal{O}(\epsilon_*^4)\Bigg\}.
\eqn
In order to work out the integral of $\sqrt{g(y)}$, we also need to specify all the expressions of $y_0(\eta),\;b_1(\eta),\;b_2(\eta),\epsilon_*(\eta)$, etc. As we have explained in Sec. III, these quantities can be expanded as follows
\bqn\lb{logexpansion}
y_0(\eta)&\simeq& \bar{y}_0+y_0'(\eta_0)\ln{\left(\frac{y}{\bar{y}_0}\right)}+\frac{1}{2}y_0''(\eta_0) \ln^2{\left(\frac{y}{\bar{y}_0}\right)},\nb\\
b_1(\eta) &\simeq & b_1(\eta_0)+b_1'(\eta_0)\ln{\left(\frac{y}{\bar{y}_0}\right)}+\frac{1}{2}b_1''(\eta_0) \ln^2{\left(\frac{y}{\bar{y}_0}\right)},\nb\\
b_2(\eta) &\simeq & b_2(\eta_0)+b_2'(\eta_0)\ln{\left(\frac{y}{\bar{y}_0}\right)}+\frac{1}{2}b_2''(\eta_0) \ln^2{\left(\frac{y}{\bar{y}_0}\right)},\nb\\
\epsilon_*(\eta) &\simeq & \epsilon_*(\eta_0)+\epsilon_*'(\eta_0)\ln{\left(\frac{y}{\bar{y}_0}\right)}+\frac{1}{2}\epsilon_*''(\eta_0) \ln^2{\left(\frac{y}{\bar{y}_0}\right)},\nb\\
\eqn
where a prime  denotes the derivative with respect to $\ln(-\eta)$.
With the above expansions, $\sqrt{g(y)}$ can be divided into three parts, corresponding to the above three expanding orders. More specifically, we have
\bqn
\sqrt{g(y)} \simeq J_0(y)+J_1(y)+J_2(y),
\eqn
where $J_0(y),\;J_1(y),\;\text{and}\;J_2(y)$ correspond to the contributions from the zeroth, first, and second-order of the above expansion. Introducing a new variable $x\equiv y/\bar{y}_0$, we have
\bqn
J_0(y)&\equiv& \sqrt{1-x^2} \left[\frac{A_1}{x}+A_2 x+A_3 x^3\right],\nb\\
J_1(y) &\equiv&  \frac{\ln x}{x \sqrt{1-x^2}}\left(B_1 +B_2 x^2+B_3 x^4+B_4 x^6 \right),\nb\\
J_2(y) &\equiv & \frac{ \ln^2x}{x \left(1-x^2\right)^{3/2}} \left(C_1 +C_2 x^2+C_3 x^4+C_4 x^6+C_5 x^8 \right),\nb\\
\eqn
here
\bqn
A_1&=&1-\frac{1}{2} \bar{b}_1 \bar y_0^2 \bar \epsilon_*^2-\frac{1}{8} \bar b_1^2 \bar{y}_0^4 \bar \epsilon _*^4+\frac{1}{2} \bar b_2 \bar y_0^4 \bar \epsilon_*^4,\nb\\
A_2&=&-\frac{1}{4} \bar b_1^2 \bar y_0^4 \bar \epsilon_*^4-\frac{1}{2} \bar b_1 \bar y_0^2 \bar \epsilon_*^2+\frac{1}{2} \bar b_2 \bar y_0^4 \bar \epsilon_*^4,\nb\\
A_3&=&\frac{1}{2} \bar b_2 \bar y_0^4 \bar \epsilon_*^4-\frac{1}{8} \bar b_1^2 \bar y_0^4 \bar \epsilon_*^4,
\eqn
\bqn
B_1&=&\frac{\bar y_0'}{\bar y_0}+\bar \epsilon_*^2 \left(-\bar b_1 \bar H_1 \bar y_0^2-\frac{\bar b_1' \bar y_0^2}{2}-\frac{3}{2} \bar b_1 \bar y_0 \bar y_0'\right)\nb\\
&&+\frac{\bar \epsilon_*^4}{8}  \Big(-4 \bar b_1^2 \bar H_1 \bar y_0^4-2 \bar b_1 \bar b_1' \bar y_0^4-5 \bar b_1^2\bar y_0^3 \bar y_0'\nb\\
&&~~~~~~~+16 \bar b_2 \bar H_1 \bar y_0^4+4 \bar b_2' \bar y_0^4+20 \bar b_2 \bar y_0^3 \bar y_0'\Big),\nb\\
B_2&=& \frac{1}{2} \bar b_1 \bar y_0 \bar \epsilon _*^2 \bar y_0'\nb\\
&&-\frac{\bar\epsilon _*^4}{4}  \left(2 \bar b_1^2 \bar H_1 \bar y_0^4+\bar b_1 \bar y_0^4 \bar b_1'+\bar b_1^2 \bar y_0^3 \bar y_0'+2 \bar b_2 \bar y_0^3 \bar y_0'\right),\nb\\
B_3&=&\bar \epsilon _*^2 \left(\bar b_1 \bar H_1 \bar y_0^2+\frac{1}{2} \bar y_0^2 \bar b_1'\right)\nb\\
&&+\frac{\bar \epsilon _*^4}{8}  \left(4 \bar b_1^2 \bar H_1 \bar y_0^4+2 \bar b_1\bar  y_0^4 \bar b_1'+3 \bar b_1^2 \bar y_0^3 \bar y_0'-4 \bar b_2 \bar y_0^3 \bar y_0'\right),\nb\\
B_4 &=&\frac{\bar  \epsilon _*^4}{4} \left(2 \bar b_1^2 \bar H_1 \bar y_0^4-8 \bar b_2 \bar H_1 \bar y_0^4+\bar b_1 \bar y_0^4 \bar b_1'-2 \bar b_2' \bar y_0^4\right),\nb\\
\eqn
and
\bqn
C_1&=&\frac{\bar y_0''}{2 \bar y_0}+\frac{ \bar \epsilon _*^2}{4} \Big(-4 \bar H_1 \bar y_0^2 \bar b_1'-12 \bar b_1 \bar H_1 \bar y_0 \bar y_0'-2 \bar b_1 \bar H_1^2 \bar y_0^2\nb\\
&&\;\;\;\;\;\;\;~~~~~~~-2 \bar b_1 \bar H_2 \bar y_0^2-6 \bar y_0 \bar b_1' \bar y_0'-6 \bar b_1 \bar y_0'^2\nb\\
&&\;\;\;\;\;\;\;\;~~~~~~~-\bar y_0^2 \bar b_1''-3\bar  b_1 \bar y_0 \bar y_0''\Big),\nb\\
 C_2&=&\frac{-\bar y_0'^2-\bar y_0 \bar y_0''}{2 \bar y_0^2}\nb\\
&&+\frac{\bar \epsilon _*^2}{4}  \Big(4 \bar H_1 \bar y_0^2 \bar b_1'+16\bar  b_1 \bar H_1 \bar y_0 \bar y_0'+2\bar  b_1 \bar H_1^2 \bar y_0^2\nb\\
&&~~~~~~~~+2 \bar b_1 \bar H_2 \bar y_0^2+8 \bar y_0 \bar b_1' \bar y_0'+9 \bar b_1 \bar y_0'^2\nb\\
&&~~~~~~~~+\bar y_0^2 \bar b_1''+4 \bar b_1 \bar y_0 \bar y_0''\Big),\nb\\
 C_3&=& \frac{\bar \epsilon _*^2}{4}  \Big(4 \bar H_1 \bar y_0^2 \bar b_1'-4 \bar b_1 \bar H_1 \bar y_0 \bar y_0'+2 \bar b_1 \bar H_1^2 \bar y_0^2+2 \bar b_1 \bar H_2 \bar y_0^2\nb\\
&&~~~~~~~~-2\bar  y_0 \bar b_1' \bar y_0'-\bar b_1 \bar y_0'^2+\bar y_0^2 \bar b_1''-\bar b_1 \bar y_0 \bar y_0''\Big),\nb\\
 C_4 &=&\frac{\bar \epsilon _*^2}{4}  \left(-4 \bar H_1 \bar y_0^2 \bar b_1'-2 \bar b_1 \bar H_1^2 \bar y_0^2-2 \bar b_1 \bar H_2 \bar y_0^2-\bar y_0^2 \bar b_1''\right),\nb\\
 C_5 &=& \mathcal{O}(\epsilon_*^2) \mathcal{O}(\epsilon^3).
 \eqn
In the above, quantities with bars   denote the ones  evaluated at the turning point $\bar y_0$, and $H_1,\;H_2,\;\text{and}\;H_3$ are defined in Appendix C.

Correspondingly, the integral of $\sqrt{g(y)}$ can also be divided into three parts
\bqn
\int_y^{\bar{y}_0} \sqrt{g(\tilde{y})} d\tilde{y} \simeq I_0+I_2+I_3,
\eqn
and in the limit $y\to 0$, we find
\bqn
\lim_{y\to 0} I_0(y) &=&-A_1 \bar{y}_0\left(1+ \ln \frac{y}{2\bar{y}_0}\right)+\frac{5A_2+2 A_3}{15}\bar{y}_0,\nb\\
\lim_{y\to 0} I_1(y) &=& -\frac{\bar{y}_0}{2} B_1 \ln^2\frac{y}{\bar y_0}-\frac{\bar{y}_0}{24} B_1 \left(\pi ^2-3 \ln^24\right)\nb\\
&&+\bar{y}_0B_2 (\ln 2-1)+\frac{\bar{y}_0}{9} B_3 (6\ln2-5)\nb\\
&&+\frac{2\bar{y}_0}{225} B_4 (60 \ln 2-47), \nb\\
\lim_{y\to 0} I_2(y) &=& \bar y_0 \Bigg\{\Bigg(-\frac{1}{3}\ln^3 \frac{y}{\bar{y}_0}+\frac{\zeta (3)}{2}+\frac{\pi ^2-\pi^2\ln 2}{12}\nb\\
&&~~~~~~~~~~~~+\frac{\ln^3 2}{3}-\ln ^22 \Bigg)C_1\nb\\
&&+\left(\frac{\pi ^2}{12}-\ln^22\right) C_2\nb\\
&&+\left(\frac{\pi ^2}{6}-2-2 \ln^22+\ln4\right) C_3\nb\\
&&+\left[\frac{2}{27} \left(3 \pi ^2-41+42 \ln 2 -36 \ln^2 2\right)\right]C_4 \nb\\
&&+\left(\frac{4 \pi ^2}{15}-\frac{4288}{1125}-\frac{16}{5}  \ln^22+\frac{296 \ln 2}{75}\right)C_5\Bigg\}.\nb\\
\eqn

\section*{Appendix B: Error control function $\mathscr{H}(\xi)$}
\renewcommand{\theequation}{B.\arabic{equation}} \setcounter{equation}{0}

From the definition of the error control function, and after some lengthy calculations, we find
\bqn
\mathscr{H}(\xi)&=&\frac{5}{36} \left\{\int_{\bar y_0}^{\tilde{y}}\sqrt{\hat{g}(\tilde{y})}d\tilde{y}\right\}^{-1}\Big|^{y}_{\bar y_0}\nb\\
&&-\int_{\bar y_0}^{y} \left\{\frac{q}{\hat{g}}-\frac{5\hat{g}'^2}{16\hat{g}^3}+\frac{\hat{g}''}{4\hat{g}^2}\right\}\sqrt{\hat{g}}dy.\nb\\
\eqn
In order to get the explicit expression of $\mathscr{H}(\xi)$, we still use the expansions of (\ref{gexpansion}) and (\ref{logexpansion}).
Thus, similar to the expansion of $\sqrt{g(y)}$, we can divide the error control function into three parts,
\bqn
\mathscr{H}(\xi)\simeq \mathscr{H}_0(\xi)+\mathscr{H}_1(\xi)+\mathscr{H}_2(\xi),
\eqn
where $\mathscr{H}_0(\xi)$, $\mathscr{H}_1(\xi)$, and $\mathscr{H}_2(\xi)$ correspond to the zeroth, first, and second order expansion
of (\ref{logexpansion}), respectively. After some tedious calculations, we find that
\bqn
\lim_{y\to 0} \mathscr{H}_0(\xi) &\simeq&\frac{1}{6 \bar{y}_0}-\frac{7\bar b_1 }{12} \bar{y}_0 \epsilon _*^2\nb\\
&&+ \left(\frac{31 \bar b_2  }{12}-\frac{109}{48} \bar{b}_1^2\right)\bar y_0^3 \epsilon _*^4,\nb\\
\lim_{y\to 0} \mathscr{H}_1(\xi) &\simeq& -\frac{\bar y_0' (23+12 \ln 2)}{72 \bar y_0^2}\nb\\
&&+\frac{1}{144}  \big(-2 \bar b_1 \bar H_1 \bar y_0-168 \bar b_1 \bar H_1 \bar y_0 \ln 2\nb\\
&&\;\;\;\;\;\;\;\;\;\;\;\;\;\;\;\;+11 \bar b_1 \bar y_0'-84 \bar b_1 \bar y_0' \ln 2-\bar b_1' \bar y_0\nb\\
&&\;\;\;\;\;\;\;\;\;\;\;\;\;\;\;\;-84 \bar b_1' \bar y_0 \ln 2\big)\bar \epsilon _*^2\nb\\
&&+\frac{1}{576}  \Big[530 \bar b_1 \bar b_1' \bar y_0^3-2616 \bar b_1 \bar b_1' \bar y_0^3 \ln 2\nb\\
&&\;\;\;\;\;\;\;\;\;+1060 \bar b_1^2 \bar H_1 \bar y_0^3-1392 \bar b_2 \bar H_1 \bar y_0^3\nb\\
&&\;\;\;\;\;\;\;\;\;-5232 \bar b_1^2 \bar H_1 \bar y_0^3 \ln 2+5952 \bar b_2 \bar H_1 \bar y_0^3 \ln 2\nb\\
&&\;\;\;\;\;\;\;\;\;+1503 \bar b_1^2 \bar y_0^2 \bar y_0'-1860 \bar b_2 \bar y_0^2 \bar y_0'\nb\\
&&\;\;\;\;\;\;\;\;\;-3924 \bar b_1^2 \bar y_0^2 \bar y_0' \ln 2+4464 \bar b_2 \bar y_0^2 \bar y_0' \ln 2\nb\\
&&\;\;\;\;\;\;\;\;\;-348 \bar b_2' \bar y_0^3+1488 \bar b_2' \bar y_0^3 \ln 2\Big]\bar \epsilon _*^4,\nb\\
\lim_{y\to 0} \mathscr{H}_2(\xi) &\simeq& \mathcal{O}(\epsilon^3)+\mathcal{O}(\epsilon^2)\mathcal{O}(\epsilon_*^2).
\eqn

\section*{Appendix C: Slow roll expansion}
\renewcommand{\theequation}{C.\arabic{equation}} \setcounter{equation}{0}

This section presents the results of the slow roll expansions of background evolution in terms of the slow roll parameters. We first consider the expansion given in  GR.

It is useful to get the exact expression of $z''(\eta)/z(\eta)$. In the single scalar field slow roll inflation, $z(\eta)$ depends on the background equation and for the scalar perturbations, we have $z_s(\eta)\equiv a\dot{\phi}/H$, where $\phi$ represents the scalar inflaton field and  dot denotes  the derivative with respect to cosmic time $t$. For the tensor perturbations, we have $z_t(\eta)=a(\eta)$. With these definitions we
 get
\bqn
\frac{z''(\eta)}{z(\eta)}&=&2a^2H^2\left(1+\epsilon+\frac{3}{2} \delta_1+2\epsilon \delta_1+\epsilon^2+\frac{1}{2} \delta_2\right),\nb\\
\;\\
\frac{a''}{a}&=&2a^2H^2 \left(1-\frac{1}{2}\epsilon\right),
\eqn
where the slow-roll parameters are defined as
\bqn
\epsilon\equiv -\frac{\dot H}{H^2}=\frac{1}{2} \left(\frac{\dot \phi}{H}\right)^2,\;\;\;\delta_n\equiv \frac{1}{H^n\dot \phi} \frac{d^{n+1}\phi}{dt^{n+1}}.
\eqn
Then, using $\dot H=-\dot\phi^2/2$, we get
\bqn
\frac{\ddot H}{H^3}=-2 \epsilon \delta_1, \;\;\;\;\;\frac{\dddot H}{H^4}=-2 \epsilon \delta_2-2\epsilon \delta_1^2,
\eqn
and
\bqn
\frac{\dot \epsilon}{H}&=&2\delta_1 \epsilon+2\epsilon^2,\\
\frac{\dot \delta_1}{H}&=&\delta_2+\delta_1\epsilon-\delta_1^2,\\
\frac{\dot \delta_2 }{H}&=&\delta_3+2 \epsilon\delta_2-\delta_2 \delta_1.
\eqn

\subsection*{1. Expansion of the conformal time $\eta$ and Hubble parameter $H$}
We also need to expand the conformal time $\eta$ in terms of the slow-roll parameters. From the relation $\eta=\int dt/a$, we obtain,
\bqn
\eta&=&\int \frac{dt}{a}=\int \frac{da}{a^2H}\nb\\
 &=&-\frac{1}{aH} \Big(1-\frac{\dot H}{H^2}-\frac{\ddot H}{H^3}+\frac{3\dot H^2}{H^4}-\frac{\dddot H}{H^4}\nb\\
&&\;\;\;\;\;\;\;\;\;\;\;\;\;\;+\frac{10 \dot H \ddot H}{H^5}-\frac{15 \dot H^3}{H^6}\Big)\nb\\
&&-\int \Big(\frac{\ddddot H}{a^2H^6}-\frac{15\dddot H \dot H}{a^2H^7}-\frac{10\ddot H^2}{a^2H^7}\nb\\
&&\;\;\;\;\;\;\;\;\;\;\;\;\;\;\;\;+\frac{105 \dot H^2 \ddot H}{a^2H^8}-\frac{105 \dot H^4}{a^2H^9}\Big)da.\eqn
In the above expressions, we only calculated the quantities  to the third-order in the slow-roll approximation.
But in principle, we can expand $\eta$ up to any order by evaluating the integral in the last two lines of the above expression. Combine all results  obtained above, we get
\bqn
\eta&\simeq&-\frac{1}{aH} \big(1+\epsilon+2 \epsilon \delta_1+3\epsilon^2+2\epsilon\delta_2\nb\\
&&\;\;\;\;\;\;\;\;\;\;\;+2 \epsilon \delta_1^2+20 \epsilon^2 \delta_1 +15 \epsilon^3+\mathcal{O}(\epsilon^4)\big).
\eqn

For the Hubble parameter $H(\eta)$, we can expand it around the turning point $y=\bar y_0$ as
\bqn
H(\eta)&\simeq&H(\eta_0)+\frac{dH(\eta)}{d\ln(-\eta)}\Big|_{\eta_0} \ln\left(\frac{y}{\bar y_0}\right)\nb\\
&&+\frac{1}{2}\frac{d^2H(\eta)}{d\ln^2(-\eta)}\Big|_{\eta_0} \ln^2\left(\frac{y}{\bar y_0}\right)\nb\\
&&+\frac{1}{6}\frac{d^3H(\eta)}{d\ln^3(-\eta)}\Big|_{\eta_0} \ln^3\left(\frac{y}{\bar y_0}\right).
\eqn
Then, it is easy to get that
\bqn
H_1&\equiv& \frac{1}{H}\frac{dH}{d\ln(-\eta)}\simeq  \epsilon+\epsilon^2+2 \epsilon^2\delta_1+3\epsilon^3,\nb\\
H_2&\equiv &\frac{1}{H}\frac{d^2H}{d\ln^2(-\eta)} \simeq -2\epsilon\delta_1-6\epsilon^2\delta_1-\epsilon^2-4\epsilon^3,\nb\\
H_3&\equiv&\frac{1}{H}\frac{d^3H}{d\ln^3(-\eta)} \simeq 2 \epsilon\delta_1+2 \epsilon\delta_1^2+3\epsilon^3+8 \epsilon^2\delta_1.\nb\\
\eqn

\subsection*{2. Expansion of $\nu_s(\eta)$ and $\nu_t(\eta)$}

Then, with the relations $\nu_s^2(\eta)=\eta^2z_s''/z_s+1/4$ and $\nu_t^2(\eta)=\eta^2a''/a+1/4$, we  get
\bqn
\nu_s(\eta)
&\simeq& \frac{3}{2} +(2\epsilon+\delta_1)+\frac{1}{3} \left(-\delta_1^2+\delta_2+14\delta_1 \epsilon+16\epsilon^2\right)\nb\\
&&+\frac{1}{9}\big(2\delta_1^3-2\delta_1\delta_2+36 \delta_1^2\epsilon+26 \delta_2 \epsilon\nb\\
&&\;\;\;\;\;\;\;\;+287 \delta_1 \epsilon^2+206 \epsilon^3\big)+\mathcal{O}(\epsilon^4),
\eqn
and
\bqn
\nu_t(\eta)
&\simeq&\frac{3}{2}+\epsilon+\frac{1}{3} \left(8\delta_1\epsilon+11\epsilon^2\right)\nb\\
&&+\frac{1}{9}\left(24\delta_1^2 \epsilon+24 \delta_2 \epsilon+236 \delta_1 \epsilon^2+173 \epsilon^3\right)\nb\\
&&\;\;+\mathcal{O}(\epsilon^4).
\eqn

Now we consider the expansion of $\nu_s(\eta)$ and $\nu_t(\eta)$ in terms of $\ln\left({\frac{y}{\bar y_0}}\right)$, which are given by
\bqn
\nu_s(\eta)&\simeq& \nu_{s0}+\bar \nu_{s1} \ln{\left(\frac{y}{\bar y_{0}}\right)} +\frac{\nu_{s2}}{2} \ln^2{\left(\frac{y}{\bar y_{0}}\right)},\nb\\
\nu_t(\eta)&\simeq& \nu_{t0}+\nu_{t1} \ln{\left(\frac{y}{\bar y_{0}}\right)} +\frac{\nu_{t2}}{2} \ln^2{\left(\frac{y}{\bar y_{0}}\right)},
\eqn
where $\nu_{s1}\equiv (d\nu_s/d\ln(-\eta))|_{\eta_{0}}$, $\nu_{s2}\equiv (d^2\nu_s/d\ln^2(-\eta))|_{\eta_{0}}$, $\nu_{t1}
\equiv (d\nu_t/d\ln(-\eta))|_{\eta_{0}}$, and $\nu_{t2}\equiv (d^2\nu_t/d\ln^2(-\eta))|_{\eta_{0}}$.
After some tedious calculations,  we obtain
\bqn
\frac{d\nu_s(\eta)}{d\ln (-\eta)}&\simeq& -(4\epsilon^2+\delta_2+5\delta_1\epsilon-\delta_1^2)\nb\\
&&-\big(-\delta_1\delta_2+\frac{2}{3} \delta_1^3+\frac{1}{3}\delta_3+\frac{19}{3}\delta_2\epsilon\nb\\
&&\;\;\;\;\;\;\;+\frac{121}{3} \delta_1 \epsilon^2+\frac{76}{3}\epsilon^3+3\epsilon\delta_1^2\big),\\
\frac{d^2\nu_s(\eta)}{d\ln^2(- \eta)}&\simeq& 16\epsilon^3+31 \epsilon^2 \delta_1+7 \epsilon \delta_2+3\epsilon \delta_1^2+\delta_3\nb\\
&&\;\;\;\;-3\delta_1 \delta_2+2\delta_1^3,
\eqn
and
\bqn
\frac{d\nu_t(\eta)}{d\ln (-\eta)}&\simeq&-(2\delta_1\epsilon+2\epsilon^2)-\frac{74}{3}\delta_1\epsilon^2-\frac{50}{3}\epsilon^3\nb\\
&&\;-\frac{8}{3}\delta_1^2\epsilon-\frac{8}{3}\epsilon\delta_2 ,\\
\frac{d^2\nu_t(\eta)}{d\ln^2(-\eta)}&\simeq & 8\epsilon^3+14 \epsilon^2\delta_1+2\epsilon \delta_1^2 +2 \epsilon \delta_2.
\eqn
It can be also shown that
\bqn
\frac{d\epsilon}{d\ln{(-\eta)}} &\simeq& - (2\epsilon^2+2 \delta_1\epsilon+2\epsilon^3+2 \delta_1 \epsilon^2),\nb\\
\frac{d^2\epsilon}{d\ln^2(-\eta)} &\simeq& 14 \delta_1 \epsilon^2+8 \epsilon^3+2 \epsilon\delta_2+2 \delta_1\epsilon,\nb\\
\frac{d\delta_1}{d\ln (-\eta)}&\simeq& - (\delta_2+\delta_1\epsilon-\delta_1^2+\epsilon \delta_2+\delta_1\epsilon^2-\epsilon \delta_1^2),\nb\\
\frac{d^2\delta_1}{d\ln^2 (-\eta)}&\simeq&\delta_3 +3 \epsilon \delta_2-3\delta_1\delta_2+3\epsilon^2\delta_1-\epsilon \delta_1^2+2\delta_1^3,\nb\\
\frac{d\delta_2}{d\ln (-\eta)}&\simeq& -\delta_3-2\epsilon\delta_2+\delta_1\delta_2.
\eqn

\section*{Appendix D: Slow roll expansion of $y_0,\;b_1,\;b_2,\;\epsilon_*$ and their derivatives}
\renewcommand{\theequation}{D.\arabic{equation}} \setcounter{equation}{0}

 First,  we consider $\epsilon_*(\eta)$, and it is easy to find that
 \bqn
 \frac{d\epsilon_*(\eta)}{d\ln(-\eta)} &=&\frac{1}{M_*} \frac{dH(\eta)}{d\ln (-\eta)}=\epsilon_* (\eta)H_1(\eta),\\
 \frac{d^2\epsilon_*(\eta)}{d\ln^2(-\eta)} &=& \frac{1}{M_*} \frac{d^2H(\eta)}{d\ln^2(-\eta)}=\epsilon_*(\eta) H_2(\eta),\\
  \frac{d^3\epsilon_*(\eta)}{d\ln^3(-\eta)} &=& \frac{1}{M_*} \frac{d^3H(\eta)}{d\ln^3(-\eta)}=\epsilon_*(\eta) H_3(\eta).
 \eqn
 For $b_1(\eta)$ and $b_2(\eta)$ we have
 \bqn
b_1(\eta)&\simeq& \hat{b}_1 (1 -2 \epsilon-3 \epsilon ^2-4 \delta _1 \epsilon-16 \epsilon ^3\nb\\
&&\;\;\;\;\;-4 \delta _2 \epsilon-28 \delta _1 \epsilon ^2-4 \delta _1^2 \epsilon ),\nb\\
b_2(\eta)&\simeq& \hat{b}_2 (  1-4 \epsilon-2 \epsilon ^2-20 \epsilon ^3 -8 \delta _1 \epsilon\nb\\
&&\;\;\;\;\;-8 \delta _2 \epsilon-40 \delta _1 \epsilon ^2-8 \delta _1^2 \epsilon ).
\eqn
Thus, we  get
\bqn
\frac{db_1(\eta)}{d\ln(-\eta)} & \simeq & \hat b_1 ( 4 \delta _1 \epsilon +4 \epsilon ^2+28 \delta _1 \epsilon ^2+4 \delta _1^2 \epsilon \nb\\
&&\;\;\;\;\;\;+4 \delta _2 \epsilon +16 \epsilon ^3),\nb\\
\frac{d^2b_1(\eta)}{d\ln^2(-\eta)} &\simeq& -\hat b_1  \left(28 \delta _1 \epsilon ^2+4 \delta _1^2 \epsilon +4 \delta _2 \epsilon +16 \epsilon ^3\right),\nb\\
\frac{db_2(\eta)}{d\ln(-\eta)} &\simeq &\hat b_2 ( 8 \delta _1 \epsilon +8 \epsilon ^2+40 \delta _1 \epsilon ^2+8 \delta _1^2 \epsilon \nb\\
&&\;\;\;\;\;\;+8 \delta _2 \epsilon +16 \epsilon ^3),\nb\\
\frac{d^2b_1(\eta)}{d\ln^2(-\eta)} &\simeq& -\hat b_2  \left(56 \delta _1 \epsilon ^2+8 \delta _1^2 \epsilon +8 \delta _2 \epsilon +32 \epsilon ^3\right).\nb\\
\eqn
\begin{widetext}
From (\ref{gofy}) we find
\bqn
y_0\simeq \nu(\eta)+\frac{b_1\nu^3(\eta)}{2} \epsilon_*^2+\frac{1}{8} (7b_1^2-4b_2)\nu^5(\eta) \epsilon_*^4+\mathcal{O}(\epsilon_*^6).
\eqn
For the scalar perturbations, we obtain
\bqn
y^{s}_0(\eta) &\simeq&\left( \frac{3}{2}+\frac{27\hat b_1}{16}  \epsilon_*^2+\frac{243(7\hat b_1^2-4\hat b_2)}{256}  \epsilon_*^4\right)+\left((2\epsilon+\delta_1)+\frac{27 \hat b_1}{8} (\epsilon+\delta_1)\epsilon_*^2+\frac{81(7\hat b_1^2-4\hat b_2)}{128}  (4\epsilon+5\delta_1) \epsilon_*^4\right)\nb\\
&&+\Bigg(\frac{14 \delta _1 \epsilon -\delta _1^2+\delta _2+16 \epsilon ^2}{3} +\frac{9\hat b_1}{16} (20 \delta _1 \epsilon +2 \delta _1^2+2 \delta _2+15 \epsilon ^2)\epsilon_*^2\nb\\
&&~~~~~~~~~~~~~~~~~~~~~~+\frac{27 (7\hat b_1^2-4 \hat b_2) }{128}\left(54 \delta _1 \epsilon +15 \delta _1^2+5 \delta _2+31 \epsilon ^2\right) \epsilon_*^4\Bigg)\nb\\
&&+\Bigg(\frac{1}{9} (287 \delta _1 \epsilon ^2+36 \delta _1^2 \epsilon +26 \delta _2 \epsilon +2 \delta _1^3-2 \delta _1 \delta _2+206 \epsilon ^3)\Bigg).
\eqn
After some tedious calculations we also obtain
\bqn
\frac{dy^s_0(\eta)}{d\ln(-\eta)} &\simeq & \left(\frac{27 \hat b_1}{8} \epsilon _*^2+\frac{243\left(7 \hat b_1^2-4 \hat b_2\right)}{64}  \epsilon _*^4 \right)\epsilon \nb\\
&&+\Bigg((-5 \delta _1 \epsilon +\delta _1^2-\delta _2-4 \epsilon ^2)+\frac{27\hat b_1 }{8}  \left(-\delta _1 \epsilon + \delta _1^2-\delta _2+ \epsilon ^2\right)\epsilon _*^2+\frac{81\left(7\hat b_1^2-4 \hat b_2\right)}{128}  \left(7 \delta _1 \epsilon +5 \delta _1^2-5 \delta _2+14 \epsilon ^2\right) \epsilon _*^4\Bigg)\nb\\
&&+\frac{1}{3} \left(-121 \delta _1 \epsilon ^2-9 \delta _1^2 \epsilon -19 \delta _2 \epsilon -2 \delta _1^3+3 \delta _1 \delta _2-\delta _3-76 \epsilon ^3\right),
 \eqn
and
\bqn
\frac{d^2y_0^s(\eta)}{d\ln^2(-\eta)} &\simeq & \left(-\frac{27\hat b_1}{4} \delta _1 \epsilon  \epsilon _*^2 +\frac{243\left(7 \hat b_1^2-4 \hat b_2\right)}{32}    \left(\epsilon^2-\delta _1 \epsilon\right)\epsilon _*^4 \right)\nb\\
&&+(31 \delta _1 \epsilon ^2+3 \delta _1^2 \epsilon +7 \delta _2 \epsilon +2 \delta _1^3-3 \delta _1 \delta _2+\delta _3+16 \epsilon ^3).
 \eqn
For the tensor perturbation, we have
\bqn
y^{t}_0(\eta) &\simeq&\left( \frac{3}{2}++\frac{27\hat b_1}{16}  \epsilon _*^2+\frac{243\left(7 \hat b_1^2-4 \hat b_2\right)}{256}  \epsilon _*^4\right)+\left(\epsilon -\frac{81\left(7 \hat b_1^2-4 \hat b_2\right)}{128}  \epsilon  \epsilon _*^4\right)\nb\\
&&+\Bigg(\frac{1}{3} \epsilon  \left(8 \delta _1+11 \epsilon \right)+\frac{9\hat b_1}{16} \epsilon   \left(4 \delta _1+5 \epsilon \right)\epsilon _*^2+\frac{27\left(7\hat  b_1^2-4 \hat b_2\right)}{64}     \left(2 \delta _1\epsilon+3 \epsilon^2 \right)\epsilon _*^4\Bigg)\nb\\
 &&+\frac{1}{9} \epsilon  \left(236 \delta _1 \epsilon +24 \delta _1^2+24 \delta _2+173 \epsilon ^2\right).
\eqn
Finally, after some tedious calculations we get
\bqn
\frac{dy^t_0(\eta)}{d\ln(-\eta)} &\simeq & \left(\frac{27\hat b_1}{8}    \epsilon _*^2+\frac{243\left(7\hat b_1^2-4 \hat b_2\right)}{64}   \epsilon _*^4 \right)\epsilon \nb\\
&&+\Bigg(\frac{81\left(7\hat b_1^2-4 \hat b_2\right)}{64}    \left(\delta _1\epsilon+2 \epsilon^2 \right)\epsilon _*^4 +\frac{27\hat b_1 }{8}  \epsilon ^2 \epsilon _*^2-2  \left(\delta _1\epsilon+\epsilon^2 \right)\Bigg)\nb\\
&&+\left(-\frac{74 \delta _1 \epsilon ^2}{3}-\frac{8 \delta _1^2 \epsilon }{3}-\frac{8 \delta _2 \epsilon }{3}-\frac{50 \epsilon ^3}{3}\right),
 \eqn
and
\bqn
\frac{d^2y_0^t(\eta)}{d\ln^2(-\eta)} &\simeq & \left(\frac{243 \left(7\hat b_1^2-4 \hat b_2\right) }{32}  \left(\epsilon^2 -\epsilon \delta _1\right) \epsilon _*^4-\frac{27\hat b_1}{4}  \left( \delta _1 \epsilon \right) \epsilon _*^2\right).
 \eqn

\end{widetext}

\end{document}